\documentclass[10pt,twocolumn,letterpaper]{article}
\usepackage[pagenumbers]{cvpr}
\usepackage{colortbl}
\usepackage{float}
\usepackage{multirow}
\usepackage{algorithm}
\usepackage{algpseudocode}
\usepackage{microtype}

\definecolor{bestcolor}{HTML}{FFCCCC}
\definecolor{secondcolor}{HTML}{FFE5B4}
\definecolor{thirdcolor}{HTML}{FFFACD}
\newcommand{\algcomment}[1]{\Statex \hspace{\algorithmicindent}\textcolor{gray}{\footnotesize$\triangleright$~#1}}
\providecommand{\Description}[1]{}

\definecolor{cvprblue}{rgb}{0.21,0.49,0.74}
\usepackage[breaklinks,colorlinks,allcolors=cvprblue]{hyperref}
\title{QuadLink: Autoregressive Quad-Dominant Mesh Generation via Point-Relation Learning}
\newcommand{\affmark}[1]{\textsuperscript{#1}}
\author{%
  {\normalsize Yiheng Zhang\affmark{1,2}\quad Zhe Zhu\affmark{2}\quad
  Tingrui Shen\affmark{2}\quad Zhuojiang Cai\affmark{3}\quad Tianxiao Li\affmark{3}}\\[2pt]
  {\normalsize Zixing Zhao\affmark{2}\quad Qiujie Dong\affmark{4}\quad
  Zhiyang Dou\affmark{4}\quad Jiepeng Wang\affmark{4}\quad Le Wan\affmark{2}}\\[2pt]
  {\normalsize Yuwang Wang\affmark{3}\quad Wenping Wang\affmark{5}\quad
  Yuan Liu\affmark{1,\(\dagger\)}\quad Cheng Lin\affmark{6,\(\dagger\)}}\\[5pt]
  {\small \affmark{1}Hong Kong University of Science and Technology
  \qquad \affmark{2}Tencent VISVISE}\\[-1pt]
  {\small \affmark{3}Tsinghua University
  \qquad \affmark{4}The University of Hong Kong}\\[-1pt]
  {\small \affmark{5}Texas A\&M University
  \qquad \affmark{6}Macau University of Science and Technology}\\[3pt]
  {\footnotesize \affmark{\(\dagger\)}Corresponding authors
  \qquad \href{https://graphic-kiliani.github.io/QuadLink-homepage/}{Project page}}%
}
\hypersetup{
  pdftitle={QuadLink: Autoregressive Quad-Dominant Mesh Generation via Point-Relation Learning},
  pdfauthor={Yiheng Zhang, Zhe Zhu, Tingrui Shen, Zhuojiang Cai, Tianxiao Li, Zixing Zhao, Qiujie Dong, Zhiyang Dou, Jiepeng Wang, Le Wan, Yuwang Wang, Wenping Wang, Yuan Liu, Cheng Lin}
}

\makeatletter
\renewcommand{\@maketitle}{%
  \begin{center}
    {\Large\bfseries \@title\par}
    \vspace{6pt}
    \begin{tabular}[t]{c}
      \@author
    \end{tabular}
  \end{center}
}
\makeatother

\begin{document}

\twocolumn[{
\renewcommand{\twocolumn}[1][]{#1}
\maketitle
\begin{minipage}{\textwidth}
\centering
\includegraphics[width=.94\textwidth]{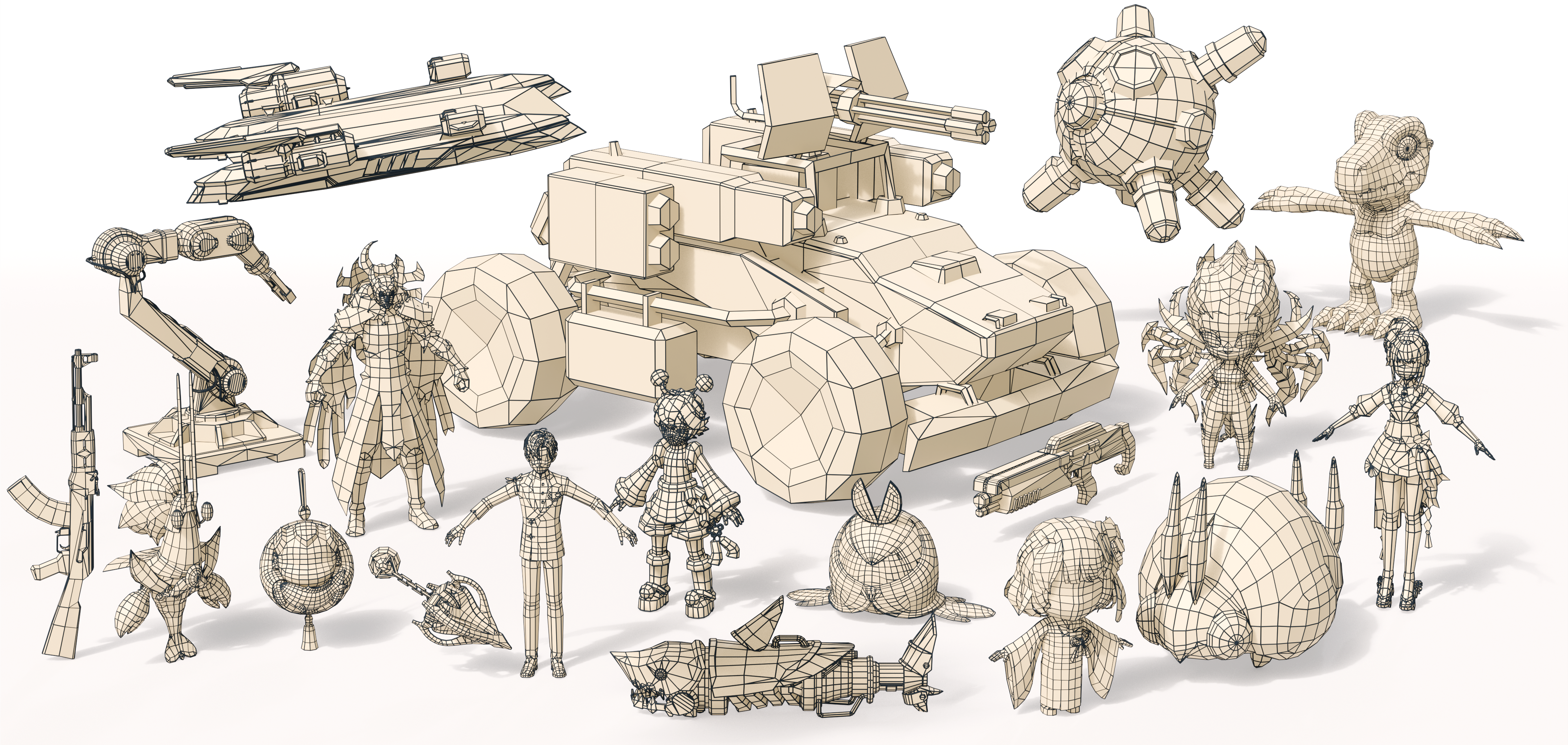}
\captionsetup{type=figure,hypcap=false}
\caption{We propose \textbf{QuadLink}, an autoregressive framework for generating production-ready quad-dominant meshes from point clouds. The combination of a face-type-agnostic point-linking architecture and high-quality anisotropic training data curated by our Tri-to-Quad Operator endows \textbf{QuadLink} with strong generalization, enabling it to generate sparse, editable meshes with coherent edge flow across diverse 3D shapes.}
\label{fig:teaser}
\end{minipage}
\par
}]

\begin{abstract}
The generation of production-ready quad-dominant meshes is a cornerstone of modern 3D content creation.
Generating anisotropic quad-dominant meshes from point clouds is challenging, as existing methods are typically limited to producing either pure triangular meshes or pure quadrilateral meshes with isotropic densities.
In this paper, we present \textbf{QuadLink}, a unified framework consisting of three stages for quad-dominant mesh generation by linking points into structured faces.
\textbf{QuadLink} formulates polygonal mesh generation as a hybrid centroid-conditioned vertex linking model: it first predicts a unified set of anchors (vertices and face centroids), then learns centroid-conditioned links that associate vertices with face centroids, and finally assembles polygonal faces with a quad-first strategy guided by robust geometric verification strategies.
This link-based formulation avoids repeatedly encoding shared vertices across faces, which enables robust and efficient generation of sparse and anisotropic quad-dominant meshes with coherent edge flow and shorter autoregressive token sequences. To construct training data for this model, we further introduce a detailed \emph{Tri-to-Quad Operator} that converts artistic triangle meshes into quad-dominant training data via geometry prefiltering and global merge selection.
Extensive experiments show that \textbf{QuadLink} produces production-ready quad-dominant meshes from point clouds and achieves improved geometric fidelity and topological quality compared to prior baselines. Meanwhile, our method natively supports hybrid polygonal topology, generalizing to arbitrary n-gon meshes without architectural changes. 

\end{abstract}

\begin{figure*}[!t]
  \centering
  \begin{minipage}[t]{0.645\textwidth}
  \vspace{0pt}
  \centering
  \includegraphics[width=\linewidth,keepaspectratio]{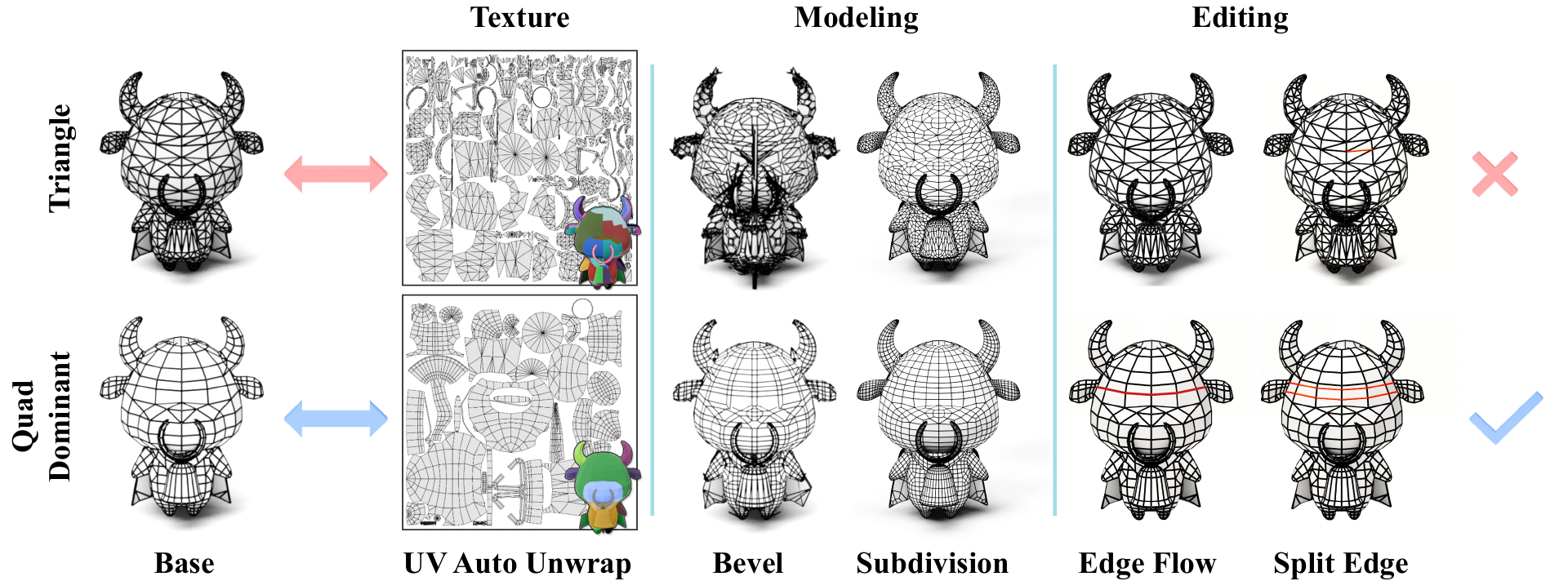}
  \captionsetup{hypcap=false}
  \captionof{figure}{\textbf{Applications of Quad-Dominant Meshes.}
  Quad-dominant meshes enable cleaner semantic UV and auto-unwrapping, support common modeling operations like beveling and subdivision, and provide coherent edge flow for controllable edge-loop editing instead of edge-by-edge editing.}
  \label{fig:application}
  \end{minipage}\hfill%
  \begin{minipage}[t]{0.333\textwidth}
  \vspace{0pt}
  \centering
  \includegraphics[width=\linewidth,keepaspectratio]{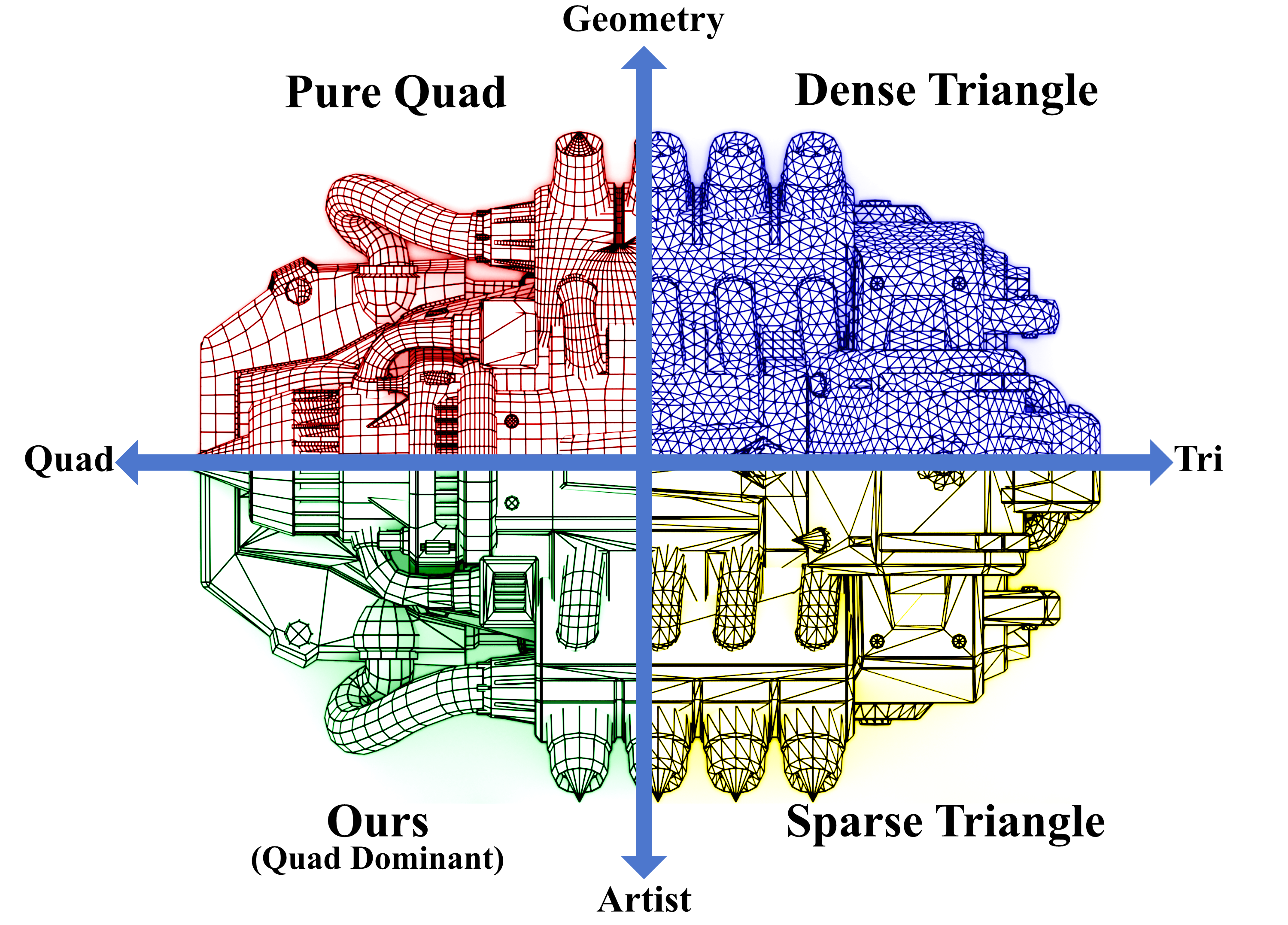}
  \captionsetup{hypcap=false}
  \captionof{figure}{
    \textbf{Artist-designed meshes differ fun-
damentally from those produced by geom-
etry processing pipelines.} We showcase their differences within the same shape.}
  \label{fig:Intro_artist_geometry}
  \end{minipage}
\end{figure*}

\section{Introduction}

Modern 3D generative models have made rapid progress in generating 3D geometry from text or images.
However, most pipelines still prioritize surface reconstruction and typically rely on implicit or volumetric representations (e.g., SDFs~\cite{zheng2022sdf}, voxels~\cite{romanelis2025efficient}, or neural fields~\cite{zhu2025neuronal}) followed by iso-surface extraction.
While effective for capturing shapes, this workflow inevitably produces dense and topologically unstructured triangle meshes, leaving the crucial problem of production-ready artistic meshes to a separate remeshing stage.
In practical content creation, topology is not merely a by-product of geometry: it directly determines whether an asset can be efficiently edited, animated, and integrated into modern production pipelines.

This gap motivates direct generation of editable mesh representations that better match production-ready topology rather than merely reconstructing surface geometry. Autoregressive mesh generation has emerged as a strong alternative by modeling meshes as discrete sequences~\cite{hao2024meshtron, zhao2025deepmesh, weng2025scaling}, demonstrating impressive capability in generating artistic triangle meshes. Yet, real production pipelines rarely stop at triangle-only outputs. Instead, they heavily rely on \emph{quad-dominant} meshes as an editable intermediate representation. A quad-dominant mesh is composed predominantly of quadrilateral faces to provide a structured and editable surface layout, while retaining a small and purposeful number of triangles to function as localized topological relaxations for accommodating geometric and topological irregularities that cannot be resolved under strict quadrilateral constraints. This structure makes quad-dominant meshes not necessarily geometry-optimal, but often production-optimal which better supports practical texture workflows and geometry operations as illustrated in Fig.~\ref{fig:application}.

Despite their practical importance, generating production-ready quad-dominant meshes remains underexplored because of three challenges. {\bfseries Hybrid Primitive Types.} Existing autoregressive methods~\cite{hao2024meshtron, weng2025scaling, zhao2025deepmesh} serialize faces, so mixed triangles and quads require padding or type-specific tokens, increasing sequence ambiguity and complicating generation. {\bfseries Anisotropic Primitive Density.} Artist-designed meshes allocate polygons non-uniformly, with large faces on simple regions and denser, directional edge flow near part boundaries and deformation axes. Such patterns are difficult to recover from dense triangulations~\cite{lee1980two, lorensen1998marching} and differ from the smooth, near-isotropic layouts favored by field-aligned remeshing methods~\cite{dong2025crossgen, tao2025learning, bommes2009mixed, knoppel2013globally}, as illustrated in Fig.~\ref{fig:Intro_artist_geometry}. {\bfseries Scarcity of Training Dataset.} Large-scale production-ready quad-dominant datasets remain limited, while common remeshing pipelines~\cite{pietroni2021reliable, huang2018quadriflow, jakob2015instant, quadremesher2019} tend to replace artistic anisotropy with geometry-driven tessellations, leaving a gap in training data.

We present {\bfseries QuadLink}, a unified framework for quad-dominant mesh generation from point clouds. To address {\bfseries hybrid primitive types}, we predict point-level anchors (vertices and face centroids) rather than serialize faces. To model {\bfseries anisotropic primitive density}, we learn centroid-conditioned vertex groups in feature space, capturing long-range, directionally structured faces. To mitigate the {\bfseries scarcity of training data}, we introduce a Tri-to-Quad Operator that turns artist-designed triangle meshes into anisotropic quad-dominant supervision. Together, these ideas support direct generation of editable meshes with mixed polygon types and coherent edge flow. 

Our main contributions are summarized as follows:
\begin{itemize}[leftmargin=1.2em,labelsep=0.4em,topsep=2pt,itemsep=1pt,parsep=0pt]
    \item \textbf{QuadLink}, a unified framework consisting of three stages for quad-dominant mesh generation with a point-centric representation for \textbf{\textit{hybrid}} polygonal layouts with substantially shorter token sequences (See analysis in Appendix Sec.~\ref{app:efficiency_token_length}.) and scalability to arbitrary n-gons.

    \item A centroid-conditioned contrastive link modeling and robust face assembly scheme that learns vertex grouping and achieves SOTA performance for \textbf{\textit{anisotropic}} production-ready topology.

    \item A SOTA tool to build high-quality quad-dominant \textit{\textbf{dataset}} with semantic anisotropy and coherent edge flow via our detailed \textit{Tri-to-Quad Operator} using geometry prefiltering and global merge selection.
\end{itemize}

\section{Related Work}

\noindent\textbf{Triangular Mesh Generation.}
Autoregressive mesh generation is a compelling paradigm for producing compact, artistic triangle meshes via causal discrete prediction.
PolyGen~\cite{nash2020polygen} pioneered the use of autoregressive models for mesh generation. Afterwards, several works explore alternative autoregressive mesh representations, including neural fields, connectivity modeling, and LLM-based text formats~\cite{shen2024spacemesh, chen2024meshxl, wang2024llama}, but with limited generation quality. MeshGPT~\cite{siddiqui2024meshgpt} established a great generative formulation that natively represents a mesh as a sequence of triangles, making autoregressive model the mainstream paradigm for artistic mesh generation. Most subsequent triangular mesh generation methods~\cite{liu2025mesh, lionar2025treemeshgpt, chen2025meshanything, tang2024edgerunner, song2025mesh, he2025charm} build upon the autoregressive paradigm and achieve competitive performance. However, encoding each triangle requires nine coordinate tokens, causing sequence length to grow linearly with face count, which severely restricts scalability and makes these methods computationally expensive for high-resolution meshes. 
As a result, subsequent research has focused on improving \emph{token compactness} to mitigate the scalability bottleneck of triangle-based autoregressive generation. Existing efforts mainly fall into several broad directions as follows.

Architectural approaches improve efficiency without changing the triangle representation, such as Meshtron~\cite{hao2024meshtron} with hierarchical factorization and hourglass transformers, and iFlame~\cite{wang2025iflame}, XSpecMesh~\cite{chen2025xspecmesh}, and FlashMesh~\cite{shen2025flashmesh} with optimized attention or decoding strategies, but they still scale linearly with face count. Direct token compression methods reduce tokens per face via blocked or hierarchical patchification~\cite{weng2025scaling,zhao2025deepmesh}, achieving substantial sequence shortening at the cost of enlarged vocabularies and degraded global connectivity. More radical representations depart from triangle sequences: FastMesh~\cite{kim2025fastmesh} generates vertex token sequence and vertex relations to greatly reduce token count, but only relies on quadratic adjacency prediction which results in chaotic connection. FACE~\cite{wang2026face} represents each triangle as a single face token for efficiency, but this abstraction reduces explicit control over vertex sharing and mesh connectivity.

Recent diffusion-based mesh generation methods~\cite{shen2024spacemesh,zhao2026lato,wang2026nexus,li2026meshflow, sun2026meshflow} explore flow matching for substantially faster mesh generation. However, latent-space decoding can limit scalability to high face counts and triangle-specific formulations do not directly accommodate mixed polygonal meshes.

\noindent \textbf{Polygonal Mesh Generation.} Directly generating quad-dominant meshes remains largely under-explored.
In practice, most quad meshes are still obtained via post-processing rather than generation.
A traditional pipeline relies on \emph{field-aligned parametrization}: it first computes a smooth cross field on the surface, then solves for a globally consistent parameterization, and extracts quadrilateral faces by tracing integer isolines or applying dedicated quad-extraction routines (e.g., Instant-Meshes~\cite{jakob2015instant}, MIQ~\cite{bommes2009mixed}, QuadriFlow~\cite{huang2018quadriflow}, QuadWild~\cite{pietroni2021reliable}, LibQEX~\cite{ebke2013qex} and recent learning-accelerated field prediction such as NeurCross~\cite{dong2024neurcross}, CrossGen~\cite{dong2025crossgen} and TopGen~\cite{chen2026topgen}). While effective for producing globally smooth and regular tessellations, these pipelines tend to over-optimize for surface smoothness and near-isotropic faces, which often deviates from modern anisotropic and semantically aligned artistic mesh layouts, and can be brittle on fine-grained details or complex topology. Some works~\cite{jiang2015frame, panozzo2014frame, corman2025rectangular, dielen2021learning} extend cross fields to anisotropic frame fields and integrable parameterizations by encoding direction-dependent scaling via metric or tensor formulations, enabling adaptive and feature-aligned quadrangulation. However, this locally defined, geometry-driven anisotropy differs from the semantically structured, non-uniform anisotropy of production-ready artist-designed meshes.

These limitations motivate quad-dominant generative formulations that can directly learn from production-ready artist-designed meshes. Point2Quad~\cite{li2025point2quad} generates quad candidates via $k$-NN grouping and filters them with an MLP classifier. However, it relies on post-processing heuristics, and the local $k$-NN formulation limits its scalability to complex geometries. QuadGPT~\cite{liu2026quadgpt} instead unifies triangles and quads by padding each face to a unified quadrilateral token length, which introduces additional padding tokens and makes face-type inference rely on sparse padding patterns, resulting in less efficient and less stable training. Mesh-Pro~\cite{zhou2026mesh} further introduces asynchronous advantage-guided ranking preference optimization to improve the quality of artist-designed quadrilateral mesh generation.

In contrast, QuadLink introduces a generative formulation for quad-dominant meshes that aligns with modern production pipelines. Unlike prior autoregressive methods that rely on triangle-only representations or padding-based unification, QuadLink directly supports structured quad-dominant topology with shorter token sequences and less information loss (See analysis in Appendix Sec.~\ref{app:efficiency_token_length}.). Together with a curated training dataset constructed via a self-developed Tri-to-Quad operator, it provides a scalable and efficient solution for quad-dominant mesh generation.

\section{QuadLink}

\begin{figure*}[t]
  \centering
  \includegraphics[width=\textwidth,keepaspectratio]{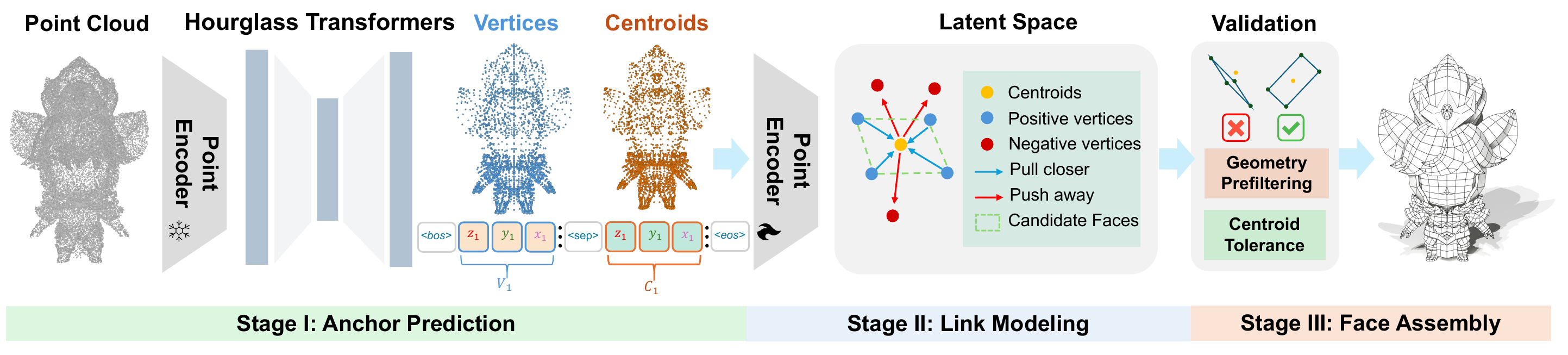}
  \caption{\textbf{Overview of QuadLink}. The pipeline consists of three stages: Stage I: Anchor Prediction, where the input point cloud is processed by a Point Cloud Encoder followed by Hourglass Transformers to generate vertex and centroid tokens. Stage II: Link Modeling, which uses contrastive learning to model the relationships between centroids and vertices. Stage III: Face Assembly, where candidate faces are progressively checked using validation criteria.}
  \label{fig:quad_baseline}
\end{figure*}

Given a surface point cloud of a 3D shape, QuadLink generates a corresponding quad-dominant mesh with high fidelity and anisotropic density. Instead of directly predicting face sequences, QuadLink adopts a centroid-conditioned vertex linking model that first generates anchors (vertices and face centroids) and then infers their connectivity to form polygonal faces. An overview of our pipeline is shown in Fig.~\ref{fig:quad_baseline}. Specifically, we first predict a compact set of anchors (Sec.~\ref{sec:I}), and then learn centroid-conditioned vertex links to capture face-level semantics (Sec.~\ref{sec:II}), and finally assemble these links into polygonal faces through a deterministic \emph{quad-first, triangle-next} assembly strategy under both the features and geometric constraints (Sec.~\ref{sec:III}).

\subsection{Stage I: Anchor Prediction}
\label{sec:I}
Given the input point cloud, the target of Stage I is to generate the vertices and face centroids, called anchors, of the underlying quad-dominant mesh. We follow an autoregressive architecture to generate these anchors as a token sequence. In the following, we first introduce how to convert these anchor points to a token sequence.

\noindent\textbf{Token Representation.}
To represent anchor points as a token sequence, we adopt the following representation:
\begingroup\setlength{\abovedisplayskip}{3pt plus 1pt minus 1pt}\setlength{\belowdisplayskip}{6pt plus 2pt minus 1pt}\setlength{\abovedisplayshortskip}{3pt plus 1pt minus 1pt}\setlength{\belowdisplayshortskip}{6pt plus 2pt minus 1pt}\begin{equation*}
\begin{aligned}
\mathcal{M}
&= \Big\{
\underbrace{z_1, y_1, x_1}_{\mathcal{T}(\mathbf{v}^{(1)})}, \ldots,
\ \tau_{\text{sep}},\
\underbrace{z'_1, y'_1, x'_1}_{\mathcal{T}(\mathbf{c}^{(1)})}, \ldots
\Big\},
\end{aligned}
\end{equation*}\endgroup
where each anchor point $(z, y, x)$ is first quantized into three discrete coordinate tokens and ordered in $z-y-x$. $\mathcal{T}$ denotes the tokenization function, and $\tau_{\mathrm{sep}}$ is a special discrete separator token that marks the start of the centroid sequence.

\noindent\textbf{Axis-aware Coordinate Tokenization.} We use a vocabulary of 3072 coordinate tokens with disjoint ranges for the three axes. The ranges are $[0,1024)$ for $z$, $[1024,2048)$ for $y$, and $[2048,3072)$ for $x$. Each token ID thus identifies both the coordinate axis and its quantized value.

\noindent\textbf{Model Architecture \& Training.}
We generate the anchor point token sequence in an autoregressive manner using the cross-entropy loss. The input point cloud is encoded with an Adaptive Michelangelo Point Cloud Encoder and the generation process is conditioned on the encoded features via cross-attention. Our generation network follows Meshtron~\cite{hao2024meshtron} to use a hierarchical transformer structure. More details are provided in Appendix Sec.~\ref{app:hourglass_arch} and~\ref{app:michelangelo_encoder}.

\subsection{Stage II: Link Modeling}
\label{sec:II}
After getting the anchor points from Anchor Prediction \textit{(Stage I)}, we proceed to link vertices into coherent polygons. To achieve this, our Link Modeling \textit{(Stage II)} aims to learn a discriminative feature space that brings incident centroid–vertex pairs closer together while pushing apart unpaired ones. This contrastive formulation encourages vertices belonging to the
same face to group around their corresponding centroid beyond Euclidean space. Consequently, each vertex can be reliably
assigned to its faces by measuring the proximity in the learned feature
space, which realizes the anisotropy of production-ready meshes. To achieve this, we finetune a Michelangelo point cloud encoder~\cite{zhao2023michelangelo}, optimized by using a contrastive-learning objective based on triplet margin loss~\cite{schroff2015facenet},  to learn such a feature space:
\begin{align}
\mathcal{L}_{\mathrm{triplet}}={}&\frac{1}{M}\sum_{i=1}^{M}\frac{1}{|\mathcal{N}^{(i)}|}\sum_{N\in\mathcal{N}^{(i)}} \Big[\|f(A^{(i)})-f(P^{(i)})\|_2^2 \notag\\&\qquad-\|f(A^{(i)})-f(N)\|_2^2+m\Big]_{+},\notag
\end{align}
\noindent where $[x]_{+}=\max(0,x)$, $A^{(i)}$ denotes the anchor (the centroid token of face $i$), 
$P^{(i)}$ is a positive vertex belonging to this face, and each $N \in \mathcal{N}^{(i)}$ denotes a negative vertex sampled from other faces.
The margin $m$ enforces a separation between positive and negative pairs in the feature space. This formulation encourages the model to generate consistent features for vertices corresponding to the same centroids, which helps connect these vertices.

\noindent\textbf{Hard Negative Mining with Adaptive Top-$K$ Selection.}
During training, we adopt a hard negative mining strategy with adaptive Top-$K$ selection. Directly optimizing the triplet loss over all negatives is often dominated by a large number of easy (near-zero) negatives, leading to weak and inefficient gradients. For each anchor--positive pair $(A^{(i)}, P^{(i)})$, we rank candidate negatives by their margin violation and select the Top-$K$ hardest ones to compute the triplet loss. Inspired by curriculum learning~\cite{bengio2009curriculum}, we employ a progressive hard-mining schedule that gradually increases $K$ during training to further stabilize optimization. Further strategy details are provided in Appendix Sec.~\ref{app:hard_mining}.

\subsection{Stage~III: Face Assembly}
\label{sec:III}
After generating the anchor points in Anchor Prediction \textit{(Stage I)} and the corresponding features in Link Modeling \textit{(Stage II)}, our Face Assembly \textit{(Stage III)} aims to connect these generated vertex points based on both the features and the geometric constraints. The key idea of Face Assembly \textit{(Stage III)} is to first search the Top-$K$ nearest vertices for each centroid in the latent feature space, and then validate whether the polygons formulated from these vertices conform to the geometric constraints. We summarize the complete procedure with pseudocode in Appendix Sec.~\ref{app:face_assembly}. Additional geometric verification criteria and thresholds are provided in Appendix Sec.~\ref{app:data_curation_details}.

\section{Quad-Dominant Data Curation}
\label{sec:data_curation}

High-quality quad-dominant mesh datasets
with production-ready topology remain extremely scarce. However,
a vast amount of artist-designed triangle meshes already exists. These
triangle meshes are typically generated by splitting the quadrilateral
faces of production-ready quad-dominant models during the asset
export or runtime stage. Therefore, if we develop an effective
reverse algorithm capable of merging adjacent triangles back into quadrilaterals while recovering artist-designed topology, we can unlock massive high-quality training data for quad-dominant mesh generation.

Appendix Sec.~\ref{app:software_remeshing_comparison} further verifies that local greedy edge merging can be suboptimal. Blossom-Quad formulates triangle pairing as a global minimum-cost perfect matching under geometric criteria, targeting pure quad meshes~\cite{remacle2012blossom}.

Our goal is instead to recover semantically structured, anisotropic quad-dominant meshes with coherent edge flow. We therefore formulate our \emph{Tri-to-Quad Operator} below.

\subsection{Global Merging Problem Formulation}\label{sec:data_problem_formulation}

Given a triangular mesh with face set $\mathcal{F}$ and candidate internal edges $E_{\mathrm{int}}$, we assign $z_e\in\{0,1\}$ to each edge $e$, where $z_e=1$ selects the merge of its two incident triangles into $q_e$. Each candidate merge receives a weight $w_e$ reflecting quad regularity and alignment with local surface flow. Score definitions and parameters are given in Appendix Sec.~\ref{app:data_curation_details}. We then solve a global maximum-weight matching problem:
\begingroup
\setlength{\abovedisplayskip}{4pt plus 1pt minus 1pt}
\setlength{\belowdisplayskip}{4pt plus 1pt minus 1pt}
\setlength{\abovedisplayshortskip}{2pt plus 1pt minus 1pt}
\setlength{\belowdisplayshortskip}{4pt plus 1pt minus 1pt}
\[\begin{gathered}
\max_{z}\; w^\top z \\
\text{s.t.}\quad Az \le \mathbf{1},\quad z_e \in \{0,1\},
\end{gathered}
\]
\endgroup
where $A$ is the face--edge incidence matrix ensuring that each triangle participates in at most one merge. More details and parameter settings are provided in Appendix Sec.~\ref{app:data_curation_details}.

\begin{figure}[t]
  \centering
  \includegraphics[width=\linewidth,keepaspectratio]{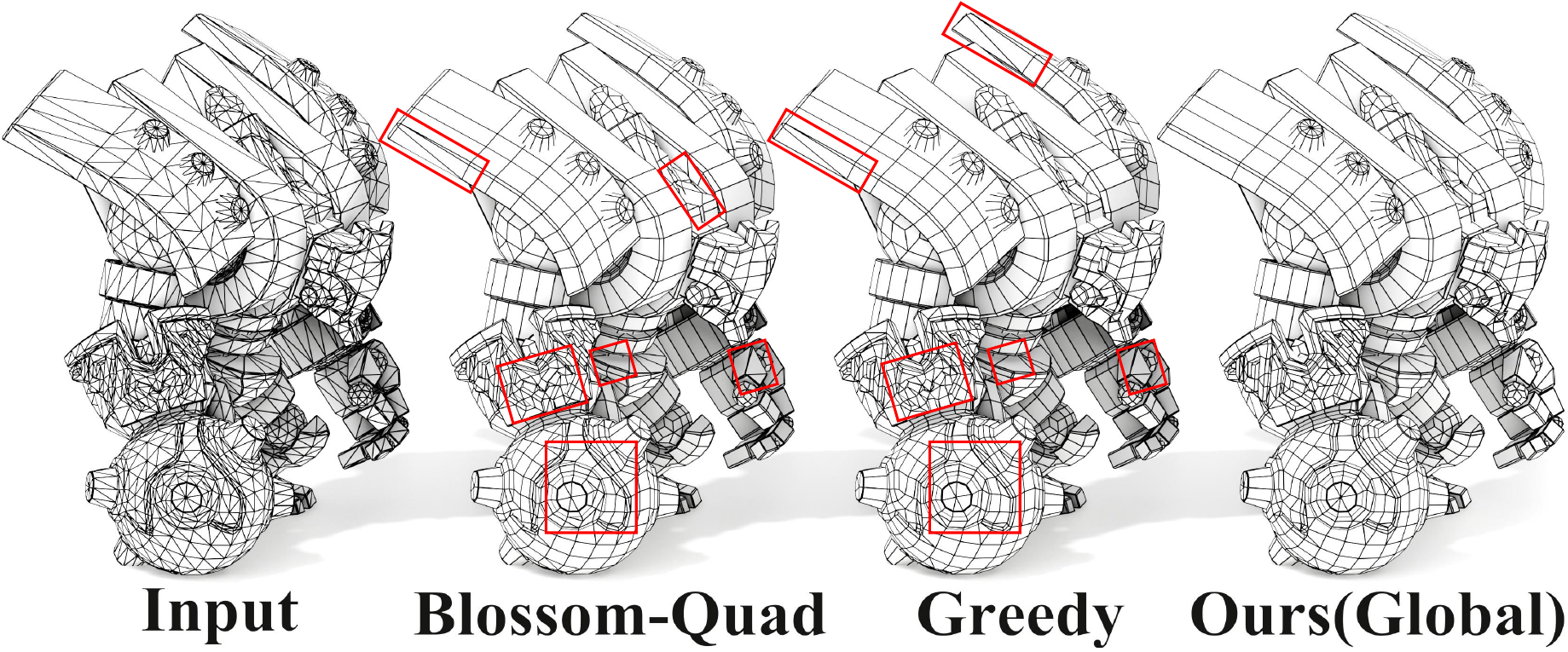}
  \captionsetup{hypcap=false}
  \caption{\textbf{Qualitative comparison with merge-edge methods.} We compare Blossom-Quad in Gmsh~\cite{geuzaine2009gmsh} with both greedy and global variants of our \emph{Tri-to-Quad Operator}. Our global formulation yields higher-quality quad-dominant meshes for data curation.}
  \label{fig:exp_operator_comparison}
\end{figure}

Unlike perfect matching, our relaxed constraint $Az\le\mathbf{1}$ allows some triangles to remain unmerged, producing quad-dominant meshes. In Fig.~\ref{fig:exp_operator_comparison}, we compare our greedy and global variants with Blossom-Quad~\cite{geuzaine2009gmsh}.

\subsection{Solving with Geometry Prefiltering}\label{sec:data_geometric_prevalidation}

Before matching, we discard candidates whose implied quads fail geometric constraints. We use the same geometric checks as in Face Assembly \textit{(Stage III)} before solving the reduced maximum-weight matching and enforce deterministic normal consistency. See more details in Appendix Sec.~\ref{app:data_curation_details}.

\section{Experiments}

\subsection{Dataset}
Both QuadLink Anchor Prediction \textit{(Stage I)} and Link Modeling \textit{(Stage II)} are trained on a curated dataset of 400k quad-dominant models, sourced from filtered internal assets licensed from 3D content providers.
Each sample is first sanitized to remove mesh degeneracies, and then converted into a quad-dominant representation via our self-developed \emph{Tri-to-Quad Operator}.
The resulting dataset spans approximately 300 mesh categories, with each model containing up to 10k points and being normalized and discretized under a 1024-level coordinate resolution.

\subsection{Experimental Protocols}
Training and inference details are provided in Appendix Sec.~\ref{app:training_details} and~\ref{app:inference_details}.

\noindent\textbf{Baselines}. For the multi-category benchmark, we consider two types of baselines.
The first category consists of recent triangle-based autoregressive models, including BPT~\cite{weng2025scaling}, DeepMesh~\cite{zhao2025deepmesh}, MeshAnythingV2~\cite{chen2025meshanything}, FastMesh~\cite{kim2025fastmesh}, TreeMeshGPT~\cite{lionar2025treemeshgpt}, MeshMosaic~\cite{xu2025meshmosaic}.
Since these approaches are fundamentally designed to output triangular artist meshes, we apply the \emph{same} \emph{Tri-to-Quad} operator as a postprocessing step to obtain quad-dominant meshes, enabling a fair and consistent evaluation.
The second category covers traditional field-aligned quad remeshing pipelines that directly extract quad meshes from the calculated cross field, for which we include three well-established methods Instant-Meshes~\cite{jakob2015instant}, QuadriFlow~\cite{huang2018quadriflow} and QuadWild~\cite{pietroni2021reliable}.

\noindent\textbf{Metrics \& Evaluation Dataset}. We evaluate geometric fidelity using Chamfer Distance (CD), Hausdorff Distance (HD), and volumetric Intersection-over-Union (IoU) between generated meshes and ground-truth meshes. To measure polygonal topology, we report the Quadrilateral Ratio (QR), defined as the proportion of quadrilateral faces among all faces. For comparisons with traditional remeshing methods, QR is often saturated since these methods typically produce pure-quad meshes. We also report Edge Flow Ratio (EFR), the mean alignment score between ground-truth feature curves and their matched output edge chains. Details of \textbf{EFR} definitions and discussions about \textbf{Topology} are provided in Appendix Sec.~\ref{sec:efr} and~\ref{app:topology_discussion}.

For evaluation, we construct a test set of 100 models, excluding meshes used for training, from both public Objaverse~\cite{deitke2023objaverse} and artist-created meshes to assess robustness across diverse shapes. We further conduct a user study with 26 professional 3D creators. Each participant evaluates 8 sampled models across ten methods under four criteria: neatness, artistry, shape fidelity and topology quality. For each criterion, the top three methods receive 3, 2, and 1 points, respectively, while the remaining receive 0 points.

\subsection{Comparison}
\begin{table}[t]
\centering
\caption{\textbf{Quantitative comparison with mesh generation and traditional quad remeshing methods.} The best scores are emphasized in \textbf{bold}, and the second are highlighted with \underline{underline}.}
\label{tab:combined_metrics}
\small
\setlength{\tabcolsep}{3pt}
\renewcommand{\arraystretch}{1.05}
\resizebox{\linewidth}{!}{%
\begin{tabular}{l|ccccc|cccc}
\toprule
\textbf{Methods}
& \multicolumn{5}{c|}{\textbf{Metrics (100 cases)}}
& \multicolumn{4}{c}{\textbf{User Study (8 cases)}} \\
\cmidrule(lr){2-6}\cmidrule(lr){7-10}
& \textbf{CD}$\downarrow$ & \textbf{HD}$\downarrow$
& \textbf{IoU}$\uparrow$ & \textbf{QR} & \textbf{EFR}$\uparrow$
& \textbf{Neatness}$\uparrow$ & \textbf{Artistry}$\uparrow$
& \textbf{Shape}$\uparrow$ & \textbf{Topology}$\uparrow$ \\
\midrule
\multicolumn{10}{l}{\textit{Mesh generation methods}} \\
BPT & 0.052 & 0.138 & 0.922 & 68.0\% & 0.328 & 0.798 & 0.755 & 0.322 & 0.572 \\
DeepMesh & 0.076 & 0.211 & 0.886 & 77.3\% & 0.381 & 0.813 & \underline{1.038} & \underline{0.909} & \underline{1.024} \\
MeshAnythingV2 & 0.063 & 0.227 & 0.889 & 71.4\% & 0.265 & 0.115 & 0.043 & 0.144 & 0.048 \\
FastMesh & 0.036 & 0.087 & 0.954 & 5.4\% & 0.244 & 0.231 & 0.524 & 0.548 & 0.298 \\
TreeMeshGPT & 0.153 & 0.372 & 0.696 & 85.9\% & 0.201 & 0.072 & 0.058 & 0.077 & 0.067 \\
MeshMosaic & 0.062 & 0.233 & 0.924 & 47.2\% & 0.292 & 0.212 & 0.452 & 0.389 & 0.404 \\
\midrule
\multicolumn{10}{l}{\textit{Traditional quad remeshing methods}} \\
Instant-Meshes & 0.038 & 0.120 & 0.905 & 100\% & 0.403 & 0.317 & 0.207 & 0.255 & 0.197 \\
QuadriFlow & 0.077 & 0.305 & 0.695 & 100\% & 0.289 & 0.096 & 0.091 & 0.125 & 0.178 \\
QuadWild & \underline{0.022} & \underline{0.086} & \underline{0.978} & 100\% & \underline{0.547} & \underline{1.168} & 0.433 & 0.779 & 0.702 \\
\midrule
\textbf{Ours}
& \textbf{0.021} & \textbf{0.056} & \textbf{0.995} & 79.4\%
& \textbf{0.714} & \textbf{2.178} & \textbf{2.399} & \textbf{2.452} & \textbf{2.510} \\
\bottomrule
\end{tabular}}
\end{table}

\begin{figure}[t]
\centering
\includegraphics[width=\linewidth]{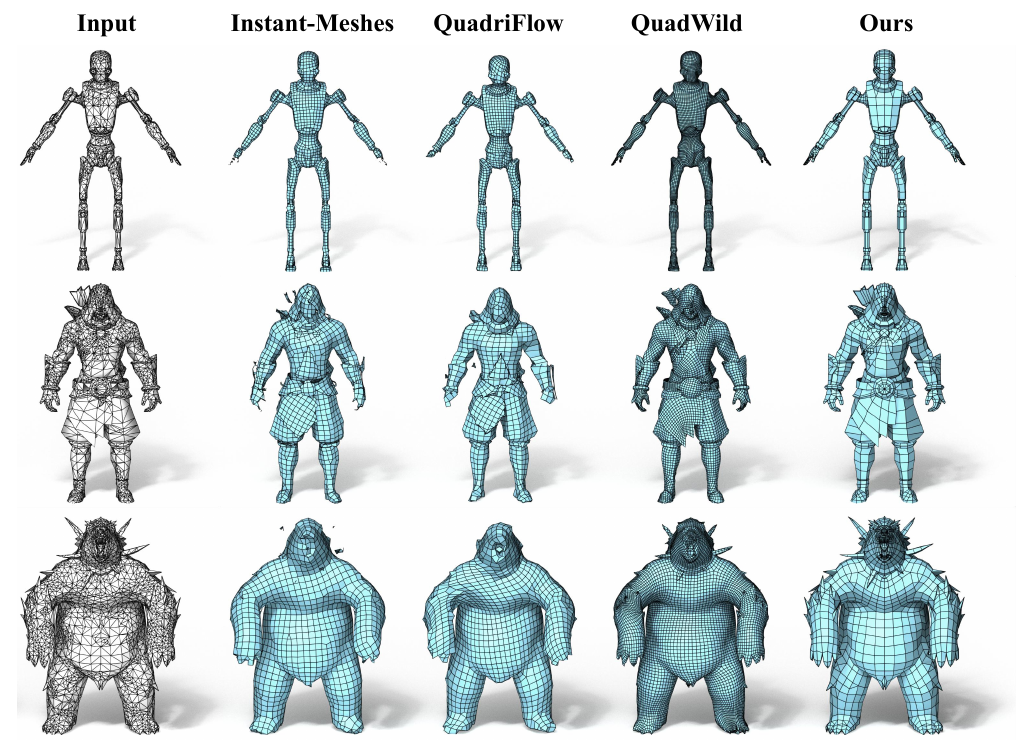}
\caption{\textbf{Qualitative comparison with field-guided remeshing methods.} They tend to produce near-isotropic layouts and are brittle on fine-grained details or complex topology.}
\label{fig:exp_comparison_quad}
\end{figure}

\begin{figure*}[!t]
\centering
\includegraphics[width=0.9\textwidth]{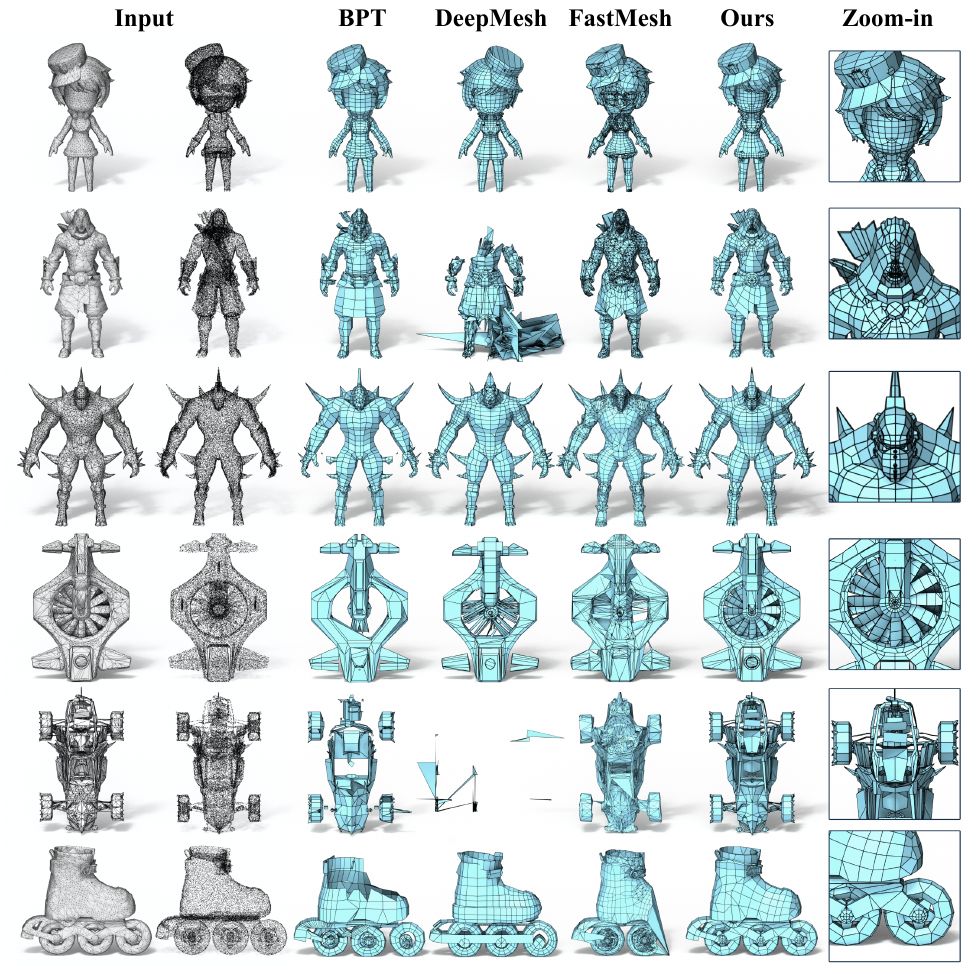}
\caption{\textbf{Qualitative comparison with triangle-based generation methods postprocessed by our \emph{Tri-to-Quad Operator}.}}
\label{fig:exp_comparison_tri}\vspace{-0.5pt}
\end{figure*}

\noindent\textbf{Comparison with Traditional Quad Remeshing Methods.} As shown in Table~\ref{tab:combined_metrics}, QuadLink outperforms traditional quad remeshing baselines across objective geometric metrics and subjective user study scores. Fig.~\ref{fig:exp_comparison_quad} further shows that field-aligned remeshing tends to produce geometry-driven, near-uniform tessellations, while QuadLink better captures the semantically structured, non-uniform anisotropy found in production-ready artist meshes. In Appendix Sec.~\ref{app:quad_remeshing_comparison}, we show how conventional quad-quality criteria can lead to unnecessarily dense faces and distort the target shape. We further verify that local decisions can lead to suboptimal merge and degraded quad quality in Appendix Sec.~\ref{app:software_remeshing_comparison}.

\noindent\textbf{Comparison with Mesh Generation Methods.} As shown in Table~\ref{tab:combined_metrics}, QuadLink surpasses all other baselines across objective geometric metrics and subjective user study scores. Qualitative comparisons in Fig.~\ref{fig:exp_comparison_tri} further show that triangle-based generative models are difficult to convert into production-ready quad-dominant meshes via geometric postprocessing. This is because their outputs are produced from discrete autoregressive coordinate sequences, where discretization errors and probabilistic decoding noise can perturb vertices that should lie on planar patches or semantically aligned edge flows. Such local inconsistencies violate the assumptions of the rule-based \emph{Tri-to-Quad operator}, making it difficult for a uniform geometric operator to robustly recover clean quad-dominant meshes. In contrast, QuadLink natively learns quad-dominant structures through centroid-conditioned vertex grouping, avoiding the fragility of triangle-first generation followed by postprocessing. More qualitative results are provided in Appendix Sec.~\ref{app:more_qualitative_comparison}.

\begin{figure}[!t]
  \centering
  \includegraphics[width=\linewidth]{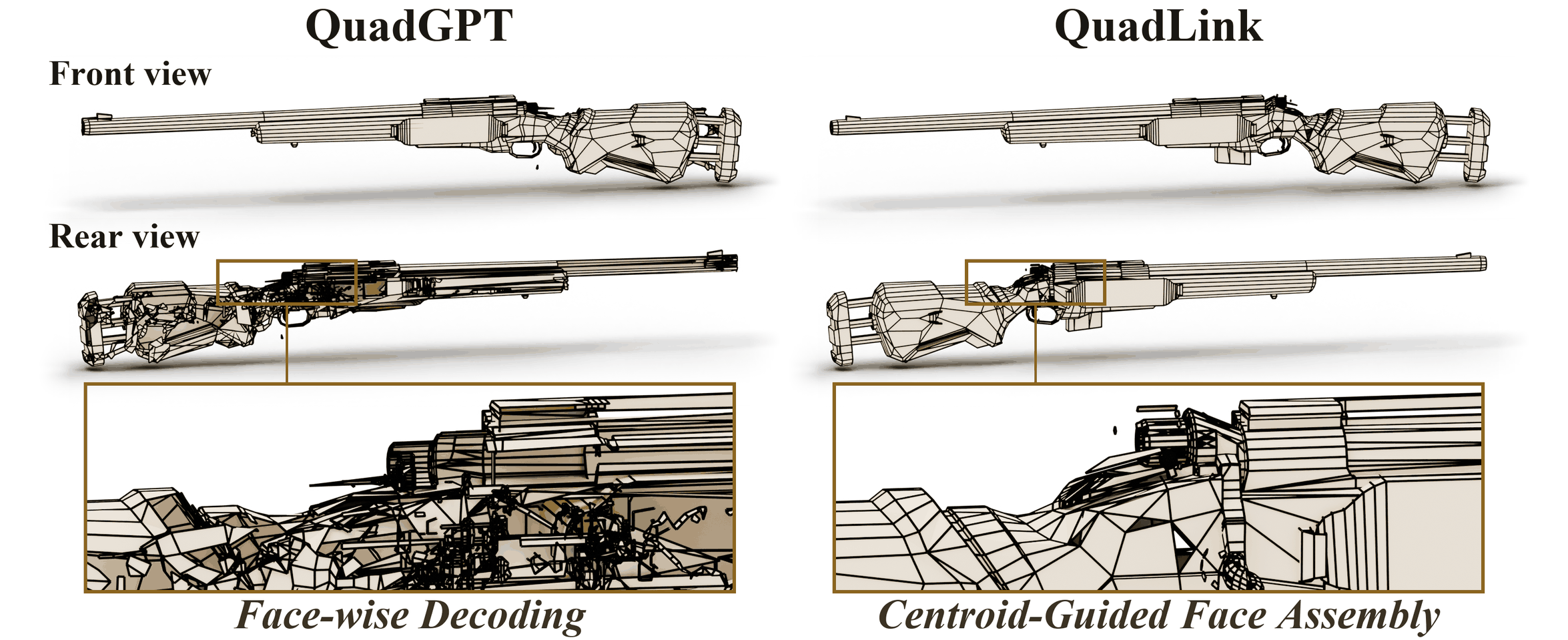}
  \captionsetup{hypcap=false}
  \caption{\textbf{Qualitative comparison with native quad-dominant mesh generation.} For QuadGPT, a token error can propagate to subsequent faces during \textit{Face-wise Decoding}, leading to fragmented and missing faces, while QuadLink preserves face connectivity outside the affected region since our \textit{Centroid-Guided Face Assembly} has shorter tokens and decouples geometry and topology.}
  \label{fig:quadgpt_face_assembly}
\end{figure}

{\clubpenalty=10000
\noindent\textbf{Comparison with Quad-Dominant Generation Methods.}
QuadGPT~\cite{liu2026quadgpt} uses 12 tokens per face, with leading padding for triangles. Mesh-Pro~\cite{zhou2026mesh} encodes three vertices before appending padding or a fourth vertex with a diagonal offset in Fig.~\ref{fig:quad_face_encoding}. Both repeat shared vertex coordinates among faces, yielding longer sequences than ours (Appendix Sec.~\ref{app:efficiency_token_length}) which can destabilize decoding and lead to ill-formed meshes or missing faces. We reimplement the supervised training stages of QuadGPT and Mesh-Pro since no public code is available and train all methods including our QuadLink from scratch on the same 3k assets from a single category. We evaluate them on 100 held-out meshes from the same category in Table~\ref{tab:gun_quad_reproduction}}. We further illustrate that QuadLink preserves face connectivity more steadily in Fig.~\ref{fig:quadgpt_face_assembly}.

\begin{figure}[!t]
\centering
\begingroup
\footnotesize
\setlength{\unitlength}{1pt}
\setlength{\fboxsep}{1.1pt}
\setlength{\tabcolsep}{3.5pt}
\newcommand{\encslot}[2]{\colorbox{#1}{\makebox[1.18em]{\strut $#2$}}}
\newcommand{\sharedslot}[1]{\encslot{cvprblue!28}{#1}}
\newcommand{\coordslot}[1]{\encslot{black!6}{#1}}
\newcommand{\padslot}{\encslot{black!14}{P}}
\raisebox{-0.5\height}{%
\begin{picture}(49,42)
  \linethickness{0.45pt}
  \qbezier(3,20)(12,14.5)(21,9)
  \qbezier(3,20)(12,26)(21,32)
  \put(21,9){\line(1,0){21}}
  \put(42,9){\line(0,1){23}}
  \put(21,32){\line(1,0){21}}
  {\color{cvprblue}\linethickness{0.9pt}%
    \put(21,9){\line(0,1){23}}
    \put(21,9){\circle*{2.6}}
    \put(21,32){\circle*{2.6}}}
  \put(3,20){\circle*{2}}
  \put(42,9){\circle*{2}}
  \put(42,32){\circle*{2}}
  \put(21,3){\makebox(0,0){\scriptsize$v_1$}}
  \put(21,38){\makebox(0,0){\scriptsize$v_2$}}
  \put(6,28){\makebox(0,0){\scriptsize$v_3$}}
  \put(42,3){\makebox(0,0){\scriptsize$v_4$}}
  \put(42,38){\makebox(0,0){\scriptsize$v_5$}}
  \put(14,20){\makebox(0,0){\scriptsize$c_T$}}
  \put(31.5,20){\makebox(0,0){\scriptsize$c_Q$}}
\end{picture}}\hspace{4pt}%
\begin{tabular}{@{}lcc@{}}
 & \textbf{Triangle} & \textbf{Quad} \\[1pt]
QuadGPT & \padslot\,\sharedslot{v_1}\,\sharedslot{v_2}\,\coordslot{v_3}
        & \sharedslot{v_1}\,\coordslot{v_4}\,\coordslot{v_5}\,\sharedslot{v_2} \\
Mesh-Pro & \sharedslot{v_1}\,\sharedslot{v_2}\,\coordslot{v_3}\,\padslot
         & \sharedslot{v_1}\,\coordslot{v_5}\,\sharedslot{v_2}\,\coordslot{v_4^\delta} \\
\noalign{\vskip 3pt\begingroup\color{black!35}\hrule height .4pt\endgroup\vskip 2pt}
\textbf{Ours} & \multicolumn{2}{c@{}}{%
  \begin{tabular}[b]{@{}c@{\enspace}c@{\enspace}c@{}}
  \textbf{Vertices} & & \makebox[0pt][c]{\textbf{Centroids}} \\[1pt]
  \sharedslot{v_1}\,\sharedslot{v_2}\,\coordslot{v_3}\,\coordslot{v_4}\,\coordslot{v_5} &
  $\tau_{\mathrm{sep}}$ &
  \encslot{green!18}{c_T}\,\encslot{green!18}{c_Q}
  \end{tabular}}
\end{tabular}
\endgroup
\captionsetup{hypcap=false}
\caption{\textbf{Shared vertices (blue) repeat in face-wise encoding.} QuadLink encodes each once, alongside the face centroids $c_T,c_Q$. $P$ denotes padding and $\delta$ the diagonal offset.}
\label{fig:quad_face_encoding}
\par\medskip
\captionof{table}{\textbf{Comparison on 100 assets with native quad-dominant generation methods.}}
\label{tab:gun_quad_reproduction}
{\small
\begin{tabular}{lcccc}
\toprule
Method & CD $\downarrow$ & HD $\downarrow$ & IoU $\uparrow$ & QR \\
\midrule
QuadGPT (reimpl.) & 0.0967 & 0.3470 & 0.3477 & 71.4\% \\
Mesh-Pro (reimpl.) & 0.0871 & 0.3283 & 0.3631 & 74.3\% \\
Ours & \textbf{0.0682} & \textbf{0.2522} & \textbf{0.4541} & 76.4\% \\
\bottomrule
\end{tabular}}
\end{figure}

\subsection{Ablation Studies}

\noindent\textbf{Anchor Prediction (Stage I).}
We ablate anchor sequence organization and axis-aware coordinate tokenization in Appendix Sec.~\ref{app:tokenization_ablation}. For the adopted anchor sequence, an axis-specific vocabulary reduces the number of epochs to convergence from 64 to 52 compared with a shared vocabulary.

\noindent\textbf{Link Modeling (Stage II).}
The learned centroid--vertex embeddings rank candidate vertices for face assembly. Appendix Sec.~\ref{app:retrieval_space_ablation} compares embedding distances, including the spacetime distance used in Nexus~\cite{wang2026nexus} and examines the effect of the supervision objective on vertex retrieval between triplet margin loss and binary cross-entropy.

\begin{figure}[!ht]
\centering
\includegraphics[width=\linewidth,height=0.21\textheight,keepaspectratio]{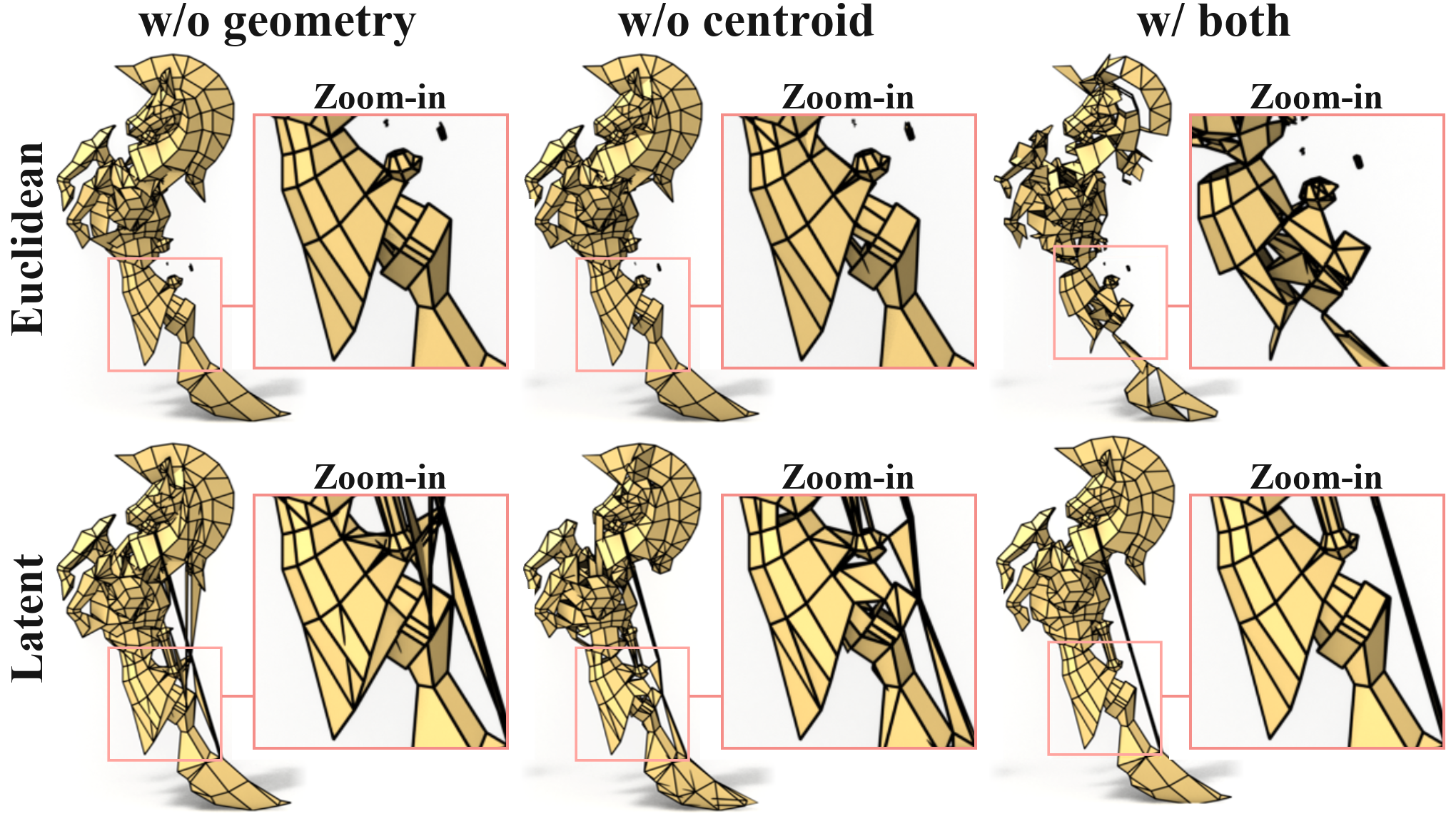}
\captionsetup{hypcap=false}
\caption{\textbf{Qualitative ablation of Face Assembly \textit{(Stage III)} under different vertex retrieval and geometric verification.}}
\label{fig:exp_stage2_ablation}
\end{figure}

\noindent\textbf{Face Assembly (Stage III).}
We compare 3D Euclidean and learned latent-space retrieval under different face verification settings in Table~\ref{tab:ablation_stage2}. With both checks enabled, latent-space retrieval lowers CD by 49.5\% and HD by 20.2\%. As shown in Fig.~\ref{fig:exp_stage2_ablation}, Euclidean retrieval fails on the seahorse reins, where nearby vertices are dominated by the dense body rather than the true incident vertices along the thin structure which requires aggressively expanding the search pool to eventually enumerate the correct face. In contrast, the learned latent space pulls incident vertices closer to their face centroid, producing compact and reliable candidate shortlists for long-range anisotropic faces. With latent-space retrieval, omitting either check reduces assembly time but raises HD from 0.0501 with both checks enabled to 0.0658 without Geometry Prefiltering and to 0.1817 without Centroid Tolerance. Geometry Prefiltering rejects geometrically invalid candidates and Centroid Tolerance checks their geometric centroids against the predicted face centroids.

\begin{table}[t]
\centering
\caption{\textbf{Quantitative ablation on Stage~III face assembly under different vertex retrieval and geometric verification settings.} The best scores are in \textbf{bold}, while the second with \underline{underline}.}
\label{tab:ablation_stage2}
\scriptsize
\setlength{\tabcolsep}{3pt}
\renewcommand{\arraystretch}{1.3}
\resizebox{\linewidth}{!}{%
\begin{tabular}{llcccc}
\toprule
\textbf{Vertex retrieval} & \textbf{Verification} &
\textbf{Time (s)}$\downarrow$ &
\textbf{Recon. rate}$\uparrow$ &
\textbf{CD}$\downarrow$ & \textbf{HD}$\downarrow$ \\
\midrule
\multirow{3}{*}{\textbf{3D Euclidean}}
& w/o geometry & 280.40 & 0.9907 & 0.0366 & 0.0746 \\
& w/o centroid tol. & \underline{176.63} & 0.9660 & 0.0392 & 0.0971 \\
& w/ both & 966.49 & 0.9549 & 0.0216 & \underline{0.0628} \\
\midrule
\multirow{3}{*}{\textbf{Learned latent}}
& w/o geometry & 258.02 & \textbf{0.9995} & \underline{0.0211} & 0.0658 \\
& w/o centroid tol. & \textbf{163.45} & 0.9667 & 0.0314 & 0.1817 \\
& w/ both (ours) & 602.64 & \underline{0.9971} & \textbf{0.0109} & \textbf{0.0501} \\
\bottomrule
\end{tabular}}
\end{table}

\noindent\textbf{Tri-to-Quad Operator (Data Curation).}
We ablate Geometry Prefiltering of candidate merges and the two merge scores for angle regularity and local surface flow alignment. Using both scores with prefiltering gives the highest OEP and EFC in Table~\ref{tab:operator_component}. We provide more details in Appendix Sec.~\ref{app:data_curation_details} for metric definitions and qualitative results.

\begin{table}[t]
\centering
\caption{\textbf{Quantitative ablation on \emph{Tri-to-Quad Operator}'s components.} The best scores are in \textbf{bold}, while the second with \underline{underline}.}
\label{tab:operator_component}
\scriptsize
\resizebox{\linewidth}{!}{%
\begin{tabular}{lcccccc}
\toprule
\multirow{2}{*}{\textbf{Metric}}
& \multicolumn{3}{c}{w/o Geometry Prefiltering}
& \multicolumn{3}{c}{w/ Geometry Prefiltering} \\
\cmidrule(lr){2-4}\cmidrule(lr){5-7}
& $w\_Q_{\mathrm{angle}}$ & $w\_Q_{\mathrm{align}}$ & w\_both
& $w\_Q_{\mathrm{angle}}$ & $w\_Q_{\mathrm{align}}$ & w\_both \\
\midrule
\textbf{OEP $\uparrow$}
& 0.9322 & 0.8248 & 0.9317 & \underline{0.9543} & 0.9439 & \textbf{0.9546} \\
\textbf{EFC $\uparrow$}
& 0.8641 & 0.7015 & 0.8642 & \underline{0.9005} & 0.8763 & \textbf{0.9014} \\
\bottomrule
\end{tabular}}
\end{table}

\section{Application}

\label{sec:production_pipeline}
Quad-dominant meshes better support practical texture workflows and geometry operations in practical \textbf{Production Pipeline} than artistic triangle meshes, like UV textures, modeling and edge-flow editing, as illustrated in Fig.~\ref{fig:application}. We also validate QuadLink's ability to generate \textbf{Polygonal Meshes} using Goldberg polyhedra~\cite{hart2012goldberg}. We provide comprehensive training details and qualitative results in Appendix Sec.~\ref{app:polygonal_mesh_generation}.

\section{Conclusion}
We presented \textbf{QuadLink}, a unified framework consisting of three stages for natively generating production-ready quad-dominant meshes. By reformulating mesh generation as a compact and centroid-conditioned vertex linking autoregressive model along with face assembly under robust geometric verification, QuadLink enables scalable polygonal generation without fragile postprocessing by introducing a robust \emph{Tri-to-Quad Operator} that provides high-quality quad-dominant supervision with structured anisotropy and coherent edge flow. Extensive experiments show that QuadLink achieves SOTA performance in all objective and subjective metrics, meanwhile making polygonal topology scalable and generating assets which are prepared for modern production pipeline. See limitations in Appendix Sec.~\ref{app:limitations}.

\section*{LLM USE STATEMENT}
We used gpt-6.1-sol to check grammar and improve the clarity of the manuscript. The research ideas originated from the authors, who also conducted the experiments and wrote the manuscript. We take responsibility for the final content of this work, including text, claims and experimental results.

\par\bigskip
\phantomsection
\pdfbookmark[0]{Appendix}{appendix}
\section*{Appendix}
\appendix

\section{Implementation Details}
\label{sec:appendix_implementation}

\subsection{Training.} For \textit{stage I Anchor Prediction}, we first freeze our adaptive Michelangelo~\cite{zhao2023michelangelo} VAE as the point cloud encoder, which encodes 49,152 surface points into 4,096 latent tokens of dimension 768. The complete VAE pipeline is applied to extract rich geometric features, which serves as cross-attention context with the interval as 1 for the mesh decoder. The decoder is a 1B-parameter autoregressive Transformer built with a two-stage hourglass architecture (depths of 8–16–8 layers), using 1,536 hidden dimensions, 16 attention heads, and a 4,096-dimension feed-forward network. To efficiently process long sequences, the model applies a linear downsampling layer with a shortening factor of 3, and adopts Rotary Position Embeddings (RoPE) with $\theta = 1\mathrm{e}{6}$ for enhanced positional encoding. We train our mesh generation model using the AdamW optimizer with $\beta_1 = 0.9$, $\beta_2 = 0.95$, a base learning rate of $1\mathrm{e}{-4}$ and a weight decay of $1\mathrm{e}{-4}$. We adopt a linear warmup followed by a cyclic cosine annealing schedule to facilitate stable and efficient convergence. 
\label{app:training_details}

For \textit{stage II Link Modeling}, we finetune a pretrained Michelangelo VAE encoder with triplet margin loss using the AdamW optimizer with a base learning rate of $1\mathrm{e}{-5}$ and a weight decay of $1\mathrm{e}{-4}$. We also employ a linear warmup followed by a cyclic cosine annealing schedule. The margin $m$ of triplet margin loss is also scheduled using a cosine warmup, starting from $0.2$ and gradually increasing to $0.3$. We adopt hard negative mining with a dynamic $k$ value that transitions from 20 to 50 as the epoch grows, combined with farthest point sampling (FPS) of 3k negative candidates per face to balance training efficiency and memory consumption.

\subsection{Inference.}
\label{app:inference_details}
For Anchor Prediction \textit{(Stage I)}, we use temperature $T = 0.5$ with top-$k$ = 20 and top-$p$ = 0.95 for sampling to balance output diversity and stability. For Link Modeling \textit{(Stage II)}, we assemble faces using the Progressive Candidate Face Selection (PCFS) strategy with a maximum pool size of $m_{max}=20$ under the latent-space ranking, and adopt a Hard-Negative Mining Schedule with $k_{start}=20$ and $k_{end}=50$. For Face Assembly \textit{(Stage III)}, we use the geometry prefiltering and centroid tolerance checks described in Appendix Section~\ref{app:data_curation_details}.

\subsection{Data Curation.}
\label{app:data_curation_details}
 For Data Curation, we convert artist-created triangle meshes into quad-dominant meshes via our \emph{Tri-to-Quad Operator} with a three-phase pipeline: geometry prefiltering, global merging and normal consistency. The geometric prefilters are also used for candidate quadrilaterals during Face Assembly \textit{(Stage III)}. Centroid consistency is evaluated in Euclidean space for both quadrilateral and triangle candidates. \paragraph{Global Merge Scoring and Selection.}
For each internal edge $e$ shared by two triangles, let $q_e$ be the resulting candidate quad, with interior angles $\{\theta_i\}_{i=1}^{4}$. We denote by $d_e$ the unit direction of $e$ and by $f_e$ the estimated local principal-curvature direction at its midpoint, computed by one-ring PCA on the tangent plane. We combine quad regularity and edge-flow alignment as:
\begin{equation}
\begin{aligned}
&Q_{\mathrm{angle}}(q_e) = \frac{1}{360^\circ}\sum_{i=1}^{4}\max\bigl(0,90^\circ-|\theta_i-90^\circ|\bigr),\\
&Q_{\mathrm{align}}(e) = \sqrt{1-(d_e\cdot f_e)^2},\\
&w_e = \alpha_1 Q_{\mathrm{angle}}(q_e)+\alpha_2 Q_{\mathrm{align}}(e).
\end{aligned}
\end{equation}
The angle score favors near-right-angle quads. The alignment score favors removing candidate edges orthogonal to the local principal direction, so the retained edges better follow the surface flow. Both scores lie in $[0,1]$; we use $\alpha_1=0.8$ and $\alpha_2=0.2$ in all experiments.

With a binary variable $z_e$ indicating whether candidate edge $e$ is merged, we select merges by solving:
\begin{equation}
\begin{gathered}[b]
\max_{z}\; w^\top z \\
\text{s.t.}\quad Az\le\mathbf{1},\quad z_e\in\{0,1\},\quad e\in E_{\mathrm{int}},
\end{gathered}
\end{equation}
where $A$ is the face--edge incidence matrix, so each triangle can participate in at most one merge. Unlike perfect matching ($Az=\mathbf{1}$), this relaxed constraint leaves some triangles unmerged and therefore supports quad-dominant output. Perfect matching instead requires every triangle to be paired, which requires an even number of triangles and a feasible matching; Blossom-Quad may otherwise rely on auxiliary edges and postprocessing heuristics.

\begin{figure*}[!t]
  \centering
  \includegraphics[width=0.95\textwidth,height=0.62\textheight,keepaspectratio]{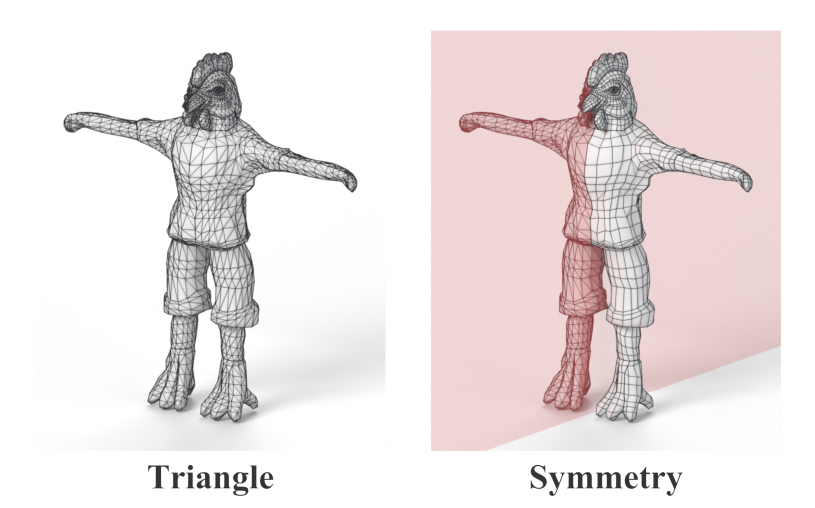}
  \captionsetup{hypcap=false,width=0.95\textwidth,margin=0pt,justification=centering,singlelinecheck=true}
  \caption{Symmetry-plane detection and edge protection in the Tri-to-Quad Operator.}
  \Description{A detected symmetry plane and the corresponding protected mesh edges.}
  \label{fig:app_operator_symmetry}
\end{figure*}

\paragraph{Geometry Prefiltering.}
Candidate quad merges are enumerated from all internal edges shared by exactly two adjacent triangles. For each candidate quadrilateral $Q$, we apply the following geometry prefiltering constraints:

\begin{enumerate}[leftmargin=1.2em, itemsep=2pt, topsep=2pt]
\item \textbf{Interior Angle constraint.}
Let $\{\theta_j\}_{j=1}^{4}$ denote the four interior angles of $Q$. To avoid extreme or degenerate quadrilaterals, we require all angles to lie within a prescribed range:
\begin{equation}
\theta_{\min} \le \theta_j \le \theta_{\max},\quad \forall j \in \{1,2,3,4\},
\end{equation}
where we set $[\theta_{\min}, \theta_{\max}] = [30^\circ, 140^\circ]$ in practice.

\item \textbf{Convexity constraint.}
We reject self-intersecting or concave quads by checking orientation consistency under a canonical vertex ordering.

\item \textbf{Dihedral constraint.}
To preserve sharp feature edges and avoid highly twisted quads, we split $Q$ along its two diagonals and require the resulting internal dihedral angles $\phi_1$ and $\phi_2$ to satisfy:
\begin{equation}
\phi_1 \le \phi_{\mathrm{thresh}},
\quad
\phi_2 \le \phi_{\mathrm{thresh}},
\end{equation}
where $\phi_{\mathrm{thresh}} = 45^\circ$.\item \textbf{Centroid constraint.}
Let $S=\{v_j\}_{j=1}^{k}$ denote the candidate-face vertices, where $k=3$ for a triangle and $k=4$ for a quadrilateral. We compute the geometric centroid in Euclidean coordinates as $\hat{c}(S)=\frac{1}{k}\sum_{j=1}^{k}v_j$, and let $c_{\mathrm{gen}}$ be the generated centroid. We accept a quadrilateral candidate only if:
\begin{equation}
\|\hat{c} - c_{\mathrm{gen}}\|_2 \le \tau_{\mathrm{quad}},
\end{equation}
where $\tau_{\mathrm{quad}} = 2 \times 10^{-3}$. For triangle candidates, we apply the same centroid validation with a separate threshold $\tau_{\mathrm{tri}} = 5 \times 10^{-3}$.

\item \textbf{UV-seam constraint.}\par
We represent the input mesh in loop-level form, where each triangle corner has its own position $p_i\in\mathbb{R}^3$ and UV coordinate $t_i\in[0,1]^2$. We set $\mathrm{weld}(i,j)=1$ if and only if the position and UV coordinates of the two corners satisfy:
\begin{equation}
\|p_i-p_j\|_\infty\le\varepsilon_{\mathrm{pos}},\quad
\|t_i-t_j\|_\infty\le\varepsilon_{\mathrm{uv}},
\label{eq:uv-weld-metric}
\end{equation}
where $\varepsilon_{\mathrm{pos}}=\varepsilon_{\mathrm{uv}}=10^{-6}$. Thus, corners at the same 3D position but on opposite sides of a UV seam retain different vertex indices. The resulting face set partitions into vertex-adjacency connected components $\{I_k\}_{k=1}^{K}$, each corresponding to one UV island. A seam edge $e$ is a boundary edge of every island it borders:
\begin{equation}
\left|\{f\in I_k:e\subset f\}\right|=1,\quad \forall I_k\text{ bordering }e.
\label{eq:uv-island-boundary}
\end{equation}
Since a merge candidate must be shared by exactly two triangles, we generate candidates independently within each island. Consequently,
\begin{equation}
E_{\mathrm{seam}}\cap E_{\mathrm{cand}}=\varnothing,
\label{eq:uv-seam-exclusion}
\end{equation}
so seam edges are never removed and the UV parametrization is preserved. Output UVs are stored per vertex in the welded index space. Fig.~\ref{fig:app_operator_uv} illustrates this seam-preserving behavior.

\begin{figure}[!t]
  \centering
  \includegraphics[width=\linewidth]{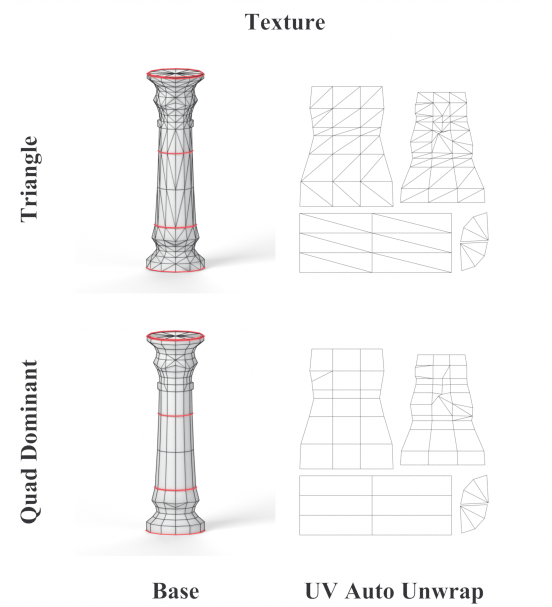}
  \captionsetup{hypcap=false,width=\linewidth,margin=0pt,justification=centering,singlelinecheck=true}
  \caption{UV-seam preservation in the Tri-to-Quad Operator.}
  \Description{UV seams are preserved during triangle-to-quad remeshing.}
  \label{fig:app_operator_uv}
\end{figure}

\item \textbf{Symmetry-plane constraint.}\par
We estimate a mirror-symmetry plane from the vertex centroid $\bar{c}$ and the eigenvectors $\{n_i\}_{i=1}^{3}$ of the vertex covariance matrix. Each principal axis defines a candidate plane $\Pi_i=\{x:n_i^\top x+d_i=0\}$ through $\bar{c}$, with $d_i=-n_i^\top\bar{c}$. We score a candidate by its reflection coverage:
\begin{equation}
\begin{aligned}
C(\Pi) &= \frac{1}{|V|}\sum_{v\in V}\mathbf{1}\!\left[\operatorname{dist}\!\left(r_\Pi(v),V\right)<\varepsilon_{\mathrm{sym}}\right],\\
r_\Pi(v) &= v-2(n^\top v+d)n,
\end{aligned}
\label{eq:symmetry-coverage}
\end{equation}
where the nearest-vertex distance is queried using a uniform grid hash and $\varepsilon_{\mathrm{sym}}=10^{-2}\,\operatorname{diag}(B)$, with $\operatorname{diag}(B)$ denoting the bounding-box diagonal. We refine the offset $d$ with a one-dimensional search and accept the plane only when $C(\Pi)\ge\tau_{\mathrm{cov}}$, where $\tau_{\mathrm{cov}}=0.9$. To reject degenerate mid-surface planes on flat or rod-like objects, we also discard a candidate if its band contains too many vertices:
\begin{equation}
\frac{\lvert\{v\in V:|n^\top v+d|<\delta\}\rvert}{|V|}>\tau_{\mathrm{band}},
\label{eq:symmetry-band-rejection}
\end{equation}
where $\tau_{\mathrm{band}}=0.2$. Given an accepted plane, an internal edge $e=(u,v)$ is protected from merging only when both endpoints lie close to that plane:
\begin{equation}
\begin{aligned}
&|n^\top u+d|<\delta_e \land |n^\top v+d|<\delta_e,\\
&\delta_e=\min\!\left(\delta,\,\beta\|u-v\|_2\right),
\end{aligned}
\label{eq:symmetry-edge-protection}
\end{equation}
with $\beta=0.2$. The global tolerance is selected adaptively, starting from a cap $\delta_{\mathrm{cap}}=5\times10^{-3}\,\operatorname{diag}(B)$. If the sorted vertex-to-plane distances contain a decisive multiplicative gap of at least $30\times$, indicating an exact on-plane vertex loop, we place $\delta$ within that gap; otherwise, we use the numerical floor $\delta=10^{-3}\,\operatorname{diag}(B)$. The edge-length-dependent term restricts protection to edges that closely follow the plane, avoiding excessive protection of thin structures pierced by it. Fig.~\ref{fig:app_operator_symmetry} shows an example of the detected plane and protected edges.
\par\smallskip

Let $E_{\mathrm{sym}}$ denote the set of protected symmetry-plane edges. The final candidate pool is:
\begin{equation}
E_{\mathrm{cand}}=E_{\mathrm{int}}\setminus\left(E_{\mathrm{seam}}\cup E_{\mathrm{sym}}\right),
\label{eq:operator-filtered-candidates}
\end{equation}
and all remaining candidates proceed through constraints 1--4 and the quality-scored matching unchanged.

\end{enumerate}

\paragraph{Global Merging.}
Details are provided in the main paper.\par\smallskip
\noindent\textbf{Tri-to-Quad Operator Ablation.}\par\smallskip

\begin{figure}[!htbp]
\centering
\includegraphics[width=\linewidth,height=0.21\textheight,keepaspectratio]{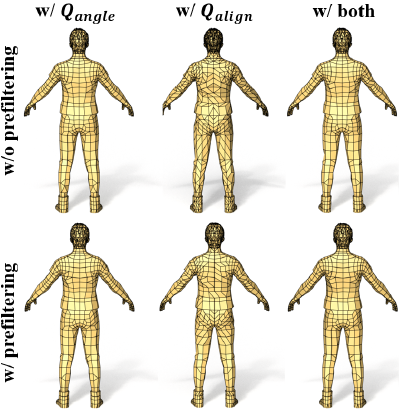}
\captionsetup{hypcap=false}
\caption{\textbf{Qualitative ablation on \emph{Tri-to-Quad Operator}.}}
\label{fig:exp_operator_ablation}
\end{figure}

\noindent
We ablate two main components in our \emph{Tri-to-Quad Operator}: Geometry Prefiltering and Merging Quality Equation $Q_{\mathrm{angle}}$ and $Q_{\mathrm{align}}$. As shown in Table~\ref{tab:operator_component} in the main paper, we apply all variants of our operator to artistic triangle meshes.
Since our merge-edge operator will preserve the original geometry with similar geometric metrics, we therefore introduce two topological metrics: \emph{Opposite Edge Parallelism} (OEP) and \emph{Edge Flow Continuity} (EFC). OEP measures the average parallelism of opposite edge pairs within each quad, with $|\cos(e_0,e_2)|$ and $|\cos(e_1,e_3)|$, reflecting local quad regularity. EFC measures the directional consistency of opposite edges across adjacent quads, reflecting edge flow smoothness. The results show that Geometry Prefiltering and $Q_{\mathrm{angle}}$ substantially improve topology by removing invalid candidates and favoring quads with regular angles. Although $Q_{\mathrm{align}}$ yields smaller quantitative gains, it is crucial in ambiguous regions with similar angle quality to produce smoother edge flow around structures such as heels, shoulders, and crotch junctions, as shown in Fig.~\ref{fig:exp_operator_ablation}.

\begin{figure}[!t]
  
  \centering
  \includegraphics[width=\linewidth,keepaspectratio]{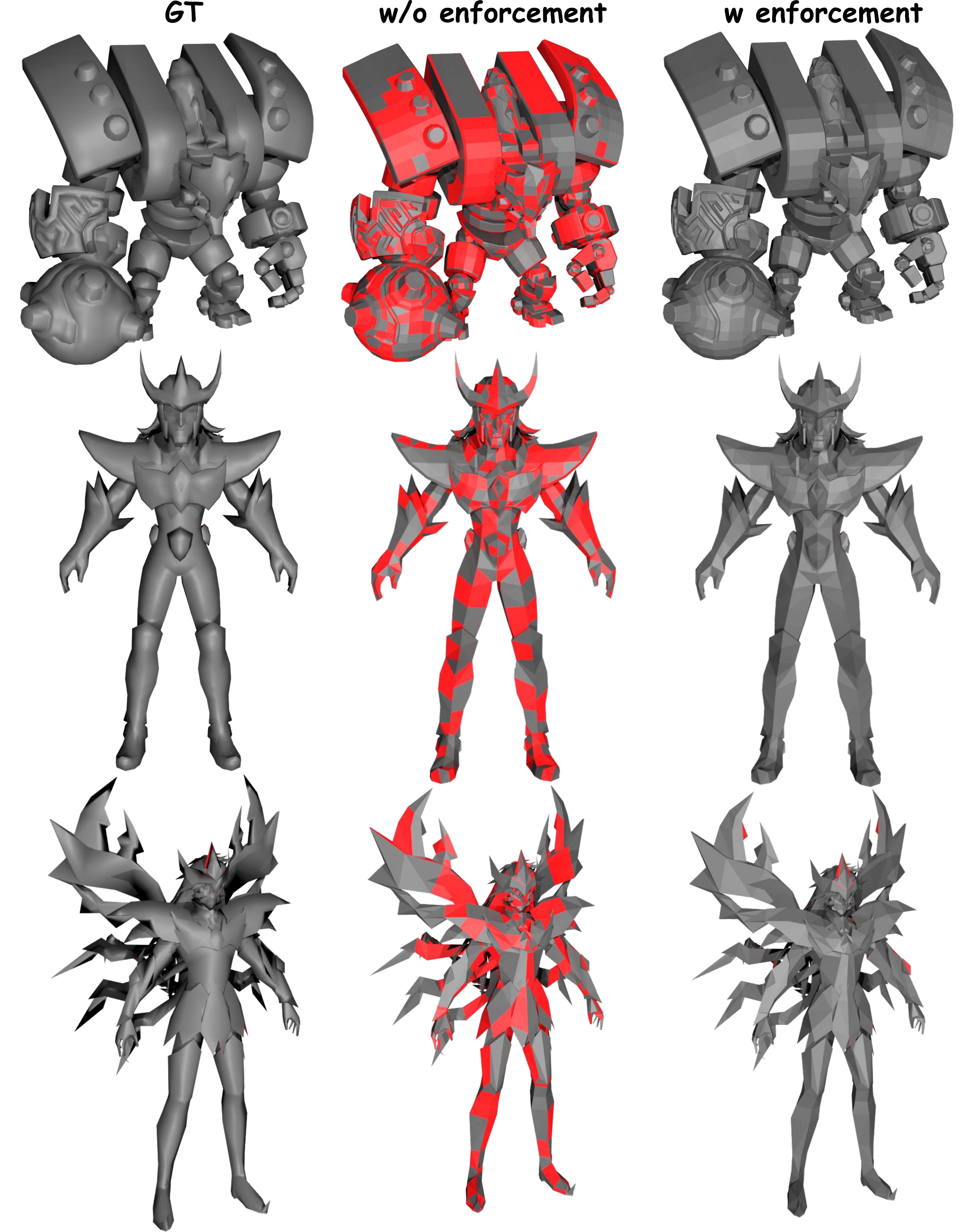}
  \caption{\textbf{Qualitative visualizations of normal consistency enforcement during triangle merging.} w/o enforcement leads to faces with inconsistent normal directions, highlighted in red (inward-facing faces). w enforcement shows results with consistent gray (outward-facing faces).}
  \Description{Normal consistency in triangle-to-quad conversion.}
  \label{fig:exp_normal_consistency}
  
\end{figure}

\paragraph{Normal Consistency.} We enforce deterministic normal consistency during merging as shown in Fig.~\ref{fig:exp_normal_consistency}: triangles are merged only if $\mathbf{n}_A\!\cdot\!\mathbf{n}_B>0$, and the resulting quad normal (computed via Newell's method) is oriented consistently by flipping vertex order if necessary, yielding globally outward-facing normals.

\section{Architecture Details}\label{sec:appendix_architecture}

\subsection{Hourglass Transformer for Stage I Anchor Prediction}
\label{app:hourglass_arch}

Rather than treating mesh token generation as a generic sequence modeling problem, we leverage the hierarchical structure of our two-level (point--coordinate) token representation and adopt the Hourglass Transformers design from Meshtron~\cite{hao2024meshtron}.
This architecture consists of a causality-preserving shortening stage followed by a symmetric upsampling stage, forming a three-block hourglass structure.
Compared to face-level autoregressive serialization, this design improves computational efficiency while retaining the essential global structure needed for Anchor Prediction \textit{Stage I} point-level generation.

Let the input token embeddings be $\mathbf{E}^{(0)} \in \mathbb{R}^{L \times D}$, where $L$ is the token length and $D$ is the embedding dimension.
The first Transformer block operates at full resolution and compresses the sequence with a shortening module using downsampling factor $r=3$, producing a compact bottleneck representation.
A second Transformer block then models dependencies at the bottleneck scale.
Finally, we restore the original temporal resolution through an upsampling stage followed by a third Transformer block:
\begin{equation}
\begin{aligned}
\mathbf{E}^{(1)} &= \mathrm{Shorten}_{3}\!\left(\mathrm{Block}_{1}\!\left(\mathbf{E}^{(0)}\right)\right)\in \mathbb{R}^{(L/3)\times D},\\
\mathbf{E}^{(2)} &= \mathrm{Block}_{2}\!\left(\mathbf{E}^{(1)}\right)\in \mathbb{R}^{(L/3)\times D},\\
\mathbf{E}^{(3)} &= \mathrm{Block}_{3}\!\left(\mathrm{Upsample}_{3}\!\left(\mathbf{E}^{(2)}\right)\right)\in \mathbb{R}^{L\times D}.
\end{aligned}
\label{eq:hourglass_3blocks_app}
\end{equation}

We follow Meshtron~\cite{hao2024meshtron} to implement $\mathrm{Shorten}_{3}(\cdot)$ and $\mathrm{Upsample}_{3}(\cdot)$ in a causality-preserving manner.
In our setting, a single shortening stage with $r=3$ is sufficient for Stage-I point-level generation, offering a favorable trade-off between efficiency and information preservation compared to padding-heavy face-level baselines.

\subsection{Adaptive Michelangelo Point Cloud Encoder}
\label{app:michelangelo_encoder}

Unlike the original Michelangelo encoder~\cite{zhao2023michelangelo}, which uses a fixed number of learnable query tokens, we additionally develop an \emph{adaptive} variant trained on multi-scale surface point clouds.
Given $N$ surface points (the union of mesh vertices and face centroids), the adaptive encoder produces a variable-length context sequence with $N$ embedding tokens, where $N$ varies across samples.
This design is mainly introduced to support Link Modeling \textit{(Stage II)}, where the number of surface points naturally changes per asset.
In Anchor Prediction \textit{(Stage I)}, we follow the canonical Michelangelo setting and retain a fixed set of learnable query tokens to obtain a compact global context for generation.

At each Transformer block, we inject the encoded point-cloud context through a residual cross-attention layer:
\begin{equation}
\mathbf{E}'=\mathbf{E}+\mathrm{CrossAttn}(\mathbf{E},\mathbf{C}),
\label{eq:crossattn_residual_app}
\end{equation}
where $\mathbf{E}\in\mathbb{R}^{L\times D}$ denotes the current token embeddings in the Hourglass Transformer, and $\mathbf{C}$ denotes the encoded point-cloud context from the Michelangelo encoder.

\subsection{Hard Negative Mining Details}
\label{app:hard_mining}
In Link Modeling \textit{(Stage II)}, we adopt Hard Negative Mining with Adaptive Top-$K$ Selection Strategy to let the model first learn coarse separation with fewer
hard negatives and later incorporate more challenging negatives
for improved discriminability.

We define the squared embedding distance as:
\begin{equation}
d(u,v)\triangleq \|f(u)-f(v)\|_2^2.
\label{eq:dist_def}
\end{equation}

For each anchor--positive pair $(A^{(i)}, P^{(i)})$, we define the margin violation of a candidate negative $N$ as
\begin{equation}
\mathrm{violation}(N)= d\!\left(A^{(i)},P^{(i)}\right)-d\!\left(A^{(i)},N\right).
\label{eq:violation}
\end{equation}

We then select the Top-$K$ negatives with the largest violation values:
\begin{equation}
\mathcal{N}^{(i)}_{\mathrm{hard}}(k)
=
\operatorname{TopK}_{N\in\mathcal{N}^{(i)}}
\big(\mathrm{violation}(N),\,k\big).
\label{eq:topk}
\end{equation}

The triplet margin loss is computed only over these hard negatives:
\begin{equation}
\begin{aligned}[b]
\mathcal{L}_{\mathrm{triplet}}^{\mathrm{hard}}
&= \frac{1}{M}\sum_{i=1}^{M}\frac{1}{|\mathcal{N}^{(i)}_{\mathrm{hard}}(k)|}\sum_{N\in\mathcal{N}^{(i)}_{\mathrm{hard}}(k)}\\
&\quad \max\bigl(0,\,d(A^{(i)},P^{(i)})-d(A^{(i)},N)+m\bigr).
\end{aligned}
\label{eq:triplet_topk}
\end{equation}

Inspired by curriculum learning~\cite{bengio2009curriculum}, we progressively increase $k$ during training:
\begin{equation}
k(t)=\min\Big(k_{\max},\ k_{\min}+\big\lfloor \alpha t \big\rfloor\Big),
\label{eq:progressive_k}
\end{equation}
where $t$ denotes the training epoch, and $k_{\min}$ and $k_{\max}$ are the initial and final Top-$K$ values.

\subsection{Stage III Face Assembly Details}
\label{app:face_assembly}

\paragraph{Contrastive Learning Trained Model (CLTM).}
\label{app:cltm}

Given a face centroid $c$ and a vertex set $\mathcal{V}$, $\texttt{CLTM}(c,\mathcal{V})$ ranks vertices by their distances to the centroid-conditioned embedding, and returns a shortlist of candidate vertices that are most likely to belong to the same face. This retrieval step significantly reduces the combinatorial search space for face reconstruction while preserving anisotropic and production-style vertex grouping learned in Link Modeling \textit{(Stage II)}.

\paragraph{Progressive Candidate Face Selection (PCFS).}\label{app:pcfs}

We reconstruct the mesh topology using a retrieve-and-verify strategy. First, for each centroid, we retrieve the Top-$K$ nearest vertices in the latent feature space as candidates. We then generate potential quadrilaterals by combining these vertices and rank them in ascending order based on the mean latent feature distance between the constituent vertices and the target centroid. We iterate through this ranked list and perform geometric verification to check for validity. We employ a \emph{quad-first, triangle-next} strategy: The first quadrilateral that satisfies these geometric constraints is selected. If no valid quadrilateral is found among the candidates, we then naturally attempt to form a valid triangle from the candidate vertices, applying the streamlined verification process to ensure topological correctness. Final centroid consistency is evaluated in Euclidean coordinates using the geometric centroid of the candidate vertices, as specified in Appendix Section~\ref{app:data_curation_details}.

We denote by $\texttt{PCFS}(c_i, P_m, k)$ a progressive enumerator that returns candidate $k$-vertex sets $S\subseteq P_m$.
During inference, we adopt a deterministic \emph{quad-first, triangle-next} assembly strategy: we first enumerate $k=4$ candidates returned by $\texttt{PCFS}(\cdot,k=4)$ and apply the validation rules; if no valid quad is found, we fall back to enumerate $k=3$ candidates via $\texttt{PCFS}(\cdot,k=3)$.
To avoid redundant computation across expanding candidate pools, combinations tested in smaller pools are cached and skipped when evaluating larger pools.

 Algorithm~\ref{alg:stage2_face_assembly} provides pseudocode for  the complete Face Assembly \textit{(Stage III)} pipeline used in our experiments. This procedure operationalizes the centroid--vertex links predicted in Link Modeling \textit{(Stage II)} by retrieving a compact candidate vertex set and progressively constructing valid polygonal faces via PCFS. The algorithm incorporates both geometric prefiltering and centroid consistency checks, enabling efficient and reliable face generation while avoiding combinatorial explosion.

\begin{algorithm*}[t]
\caption{Stage III Face Assembly with Geometry Prefiltering and Centroid Tolerance}
\label{alg:stage2_face_assembly}
\scriptsize
\begin{algorithmic}[1]
\Require Centroid set $\mathcal{C}$, vertex set $\mathcal{V}$
\Require Contrastive Learning Trained Model $\texttt{CLTM}(\cdot)$ in ~\ref{app:cltm}
\Require Progressive Candidate Face Selection $\texttt{PCFS}(\cdot)$ in ~\ref{app:pcfs}
\Require Geometry prefiltering rules $\texttt{GeoFilter}(\cdot)$
\Require Centroid tolerance $\tau_{\text{quad}}, \tau_{\text{tri}}$ 
\Ensure Constructed face set $\mathcal{F}$

\State $\mathcal{F} \gets \emptyset$
\For{each centroid $c \in \mathcal{C}$}
    \State $\mathcal{V}_{\text{cand}} \gets \texttt{TopK}\big(\texttt{CLTM}(c, \mathcal{V}), K=20\big)$
    \State $\texttt{found} \gets \textbf{false}$
    
    \algcomment{Quad-first}
    \For{each 4-vertex set $S \gets \texttt{PCFS}(c,\mathcal{V}_{\text{cand}}, k=4)$}
        \If{$\texttt{GeoFilter}(S)$ \textbf{and} $\texttt{QuadCentroidTol}(S, c) \le \tau_{\text{quad}}$}
            \State $\mathcal{F} \gets \mathcal{F} \cup \{\texttt{MakeQuad}(S)\}$
            \State $\texttt{found} \gets \textbf{true}$
            \State \textbf{break}
        \EndIf
    \EndFor
    
    \algcomment{Tri-Fallback}
    \If{\textbf{not} $\texttt{found}$}
        \For{each 3-vertex set $S \gets \texttt{PCFS}(c,\mathcal{V}_{\text{cand}}, k=3)$}
            \If{$\texttt{GeoFilter}(S)$ \textbf{and} $\texttt{TriCentroidTol}(S, c) \le \tau_{\text{tri}}$}
                \State $\mathcal{F} \gets \mathcal{F} \cup \{\texttt{MakeTri}(S)\}$
                \State \textbf{break}
            \EndIf
        \EndFor
    \EndIf
\EndFor

\State \Return $\mathcal{F}$
\end{algorithmic}
\end{algorithm*}

\subsection{Mesh Tokenization and Vocabulary Strategies}
\label{app:seq_organizations_vocab_stategies}

Anchor Prediction \textit{(Stage I)} predicts a unified set of anchors, including both mesh vertices and face centroids. This introduces two practical tokenization challenges for autoregressive sequence modeling. First, the sequence must distinguish different point types, i.e., vertices and centroids. Second, each 3D coordinate token must encode its axis identity ($z$, $y$, or $x$). We therefore study both sequence organization strategies and vocabulary designs.

\paragraph{Sequence Organization.}
To analyze how point-type organization affects convergence and training stability, we evaluate four representative tokenization modes.

\noindent\textit{Single.}
This mode serializes only vertices sorted by the $z$--$y$--$x$ lexicographic order. It serves as a simplified baseline that isolates the optimization difficulty without mixing vertices and centroids:
\begin{equation*}
\mathcal{M}
=
\Big\{
\mathbf{v}^{(1)}, \mathbf{v}^{(2)}, \ldots
\Big\}.
\end{equation*}

\noindent\textit{Dual Codebook (Mixed).}
This mode mixes vertices and centroids into a single sequence and jointly sorts them by the $z$--$y$--$x$ lexicographic order. To distinguish the two point types, vertex coordinates and centroid coordinates are encoded using disjoint vocabularies:
\begin{equation*}
\mathcal{M}
=
\Big\{
\mathbf{v}^{(1)}, \mathbf{c}^{(1)}, \mathbf{v}^{(2)}, \ldots
\Big\}.
\end{equation*}

\noindent\textit{Dual Codebook (Separate).}
This mode separates vertices and centroids into two consecutive blocks, each sorted by the $z$--$y$--$x$ lexicographic order. Similar to the mixed variant, it uses disjoint coordinate vocabularies for vertices and centroids:
\begin{equation*}
\mathcal{M}
=
\Big\{
\underbrace{\mathbf{v}^{(1)}, \ldots}_{\mathbf{V}},
\underbrace{\mathbf{c}^{(1)}, \ldots}_{\mathbf{C}}
\Big\}.
\end{equation*}

\noindent\textit{Single Separate.}
This mode also separates vertices and centroids into two blocks, but inserts an explicit \texttt{<sep>} token between them. The delimiter provides sequence-level type separation, allowing vertices and centroids to share the same coordinate vocabulary:
\begin{equation*}
\mathcal{M}
=
\Big\{
\underbrace{\mathbf{v}^{(1)}, \ldots}_{\mathbf{V}},
\texttt{<sep>},
\underbrace{\mathbf{c}^{(1)}, \ldots}_{\mathbf{C}}
\Big\}.
\end{equation*}

\paragraph{Axis-aware Vocabulary.}
Besides point-type organization, we further study whether coordinate axes should share a vocabulary. A common strategy is to use the same coordinate vocabulary for $z$, $y$, and $x$, relying on positional embedding such as RoPE~\cite{su2024roformer} to infer axis identity. In contrast, our axis-aware vocabulary assigns disjoint index ranges to different axes, so that each coordinate token explicitly encodes whether it corresponds to $z$, $y$, or $x$.

As shown in Tab.~\ref{tab:token_vocab_ablation}, axis-aware factorization consistently improves convergence across tokenization modes. We attribute this to the reduced burden of implicit classification: the model no longer needs to infer both point type and coordinate axis solely from sequence context. Among all variants, \textit{Single Separate} with axis-aware vocabulary achieves the best convergence efficiency. We therefore adopt this design in our final model.

\begin{table*}[t]
\centering
\caption{\textbf{Tokenization ablation for Stage~I anchor prediction.}
We evaluate different token structures and per-axis vocabulary factorization, reporting vocabulary size and convergence epoch (loss $\le 0.1$).
Axis-aware tokenization improves convergence across nearly all schemes. Among the variants that predict both vertices and centroids, axis-aware single\_separate reaches the loss threshold earliest.
Results are obtained on a category-balanced subset of 2,000 assets and verified to hold on the full training set.}
\label{tab:token_vocab_ablation}
\small
\setlength{\tabcolsep}{4pt}
\renewcommand{\arraystretch}{1.15}
\resizebox{\linewidth}{!}{\begin{tabular}{llp{8.2cm}cc}
\toprule
\textbf{Method} & \textbf{Axis-aware vocabulary} & \textbf{Token Structure} & \textbf{Vocab Size} & \textbf{Epoch to convergence} \\
\midrule
\multirow{2}{*}{single}
& false & vertex [0--1023] & 1024 & 55 \\
& true & $z$[0--1023], $y$[1024--2047], $x$[2048--3071] & 3072 & 36 \\
\midrule
\multirow{2}{*}{dual}
& false & vertex [0--1023], centroid [1024--2047] & 2048 & 74 \\
& true & $z_v$[0--1023], $y_v$[1024--2047], $x_v$[2048--3071], \newline
$z_c$[3072--4095], $y_c$[4096--5119], $x_c$[5120--6143] & 6144 & 66 \\
\midrule
\multirow{2}{*}{dual\_separate}
& false & vertex [0--1023], centroid [1024--2047] & 2048 & 59 \\
& true & $z_v$[0--1023], $y_v$[1024--2047], $x_v$[2048--3071], \newline
$z_c$[3072--4095], $y_c$[4096--5119], $x_c$[5120--6143] & 6144 & 65 \\
\midrule
\multirow{2}{*}{single\_separate}
& false & vertex [0--1023] + $\langle\mathrm{SEP}\rangle$ + centroid [0--1023] & 1025 & 64 \\
& true & $z$[0--1023], $y$[1024--2047], $x$[2048--3071] + $\langle\mathrm{SEP}\rangle$ & 3073 & 52 \\
\bottomrule
\end{tabular}}
\end{table*}

\subsection{Ablation on Tokenization Selection}
\label{app:tokenization_ablation}
\begin{figure}[t]
  \centering
  \includegraphics[width=\linewidth,
                   height=0.28\textheight,
                   keepaspectratio]{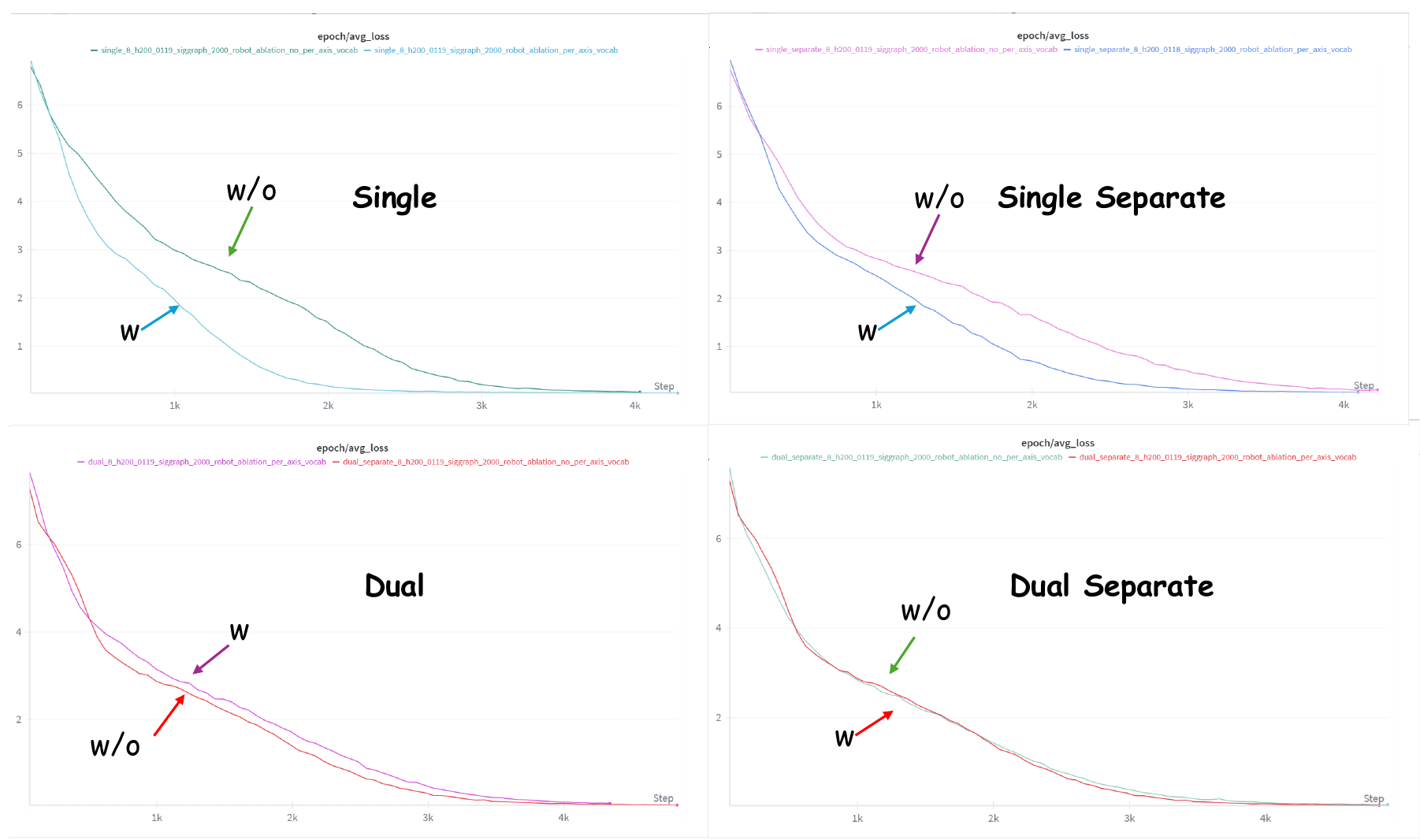}
  \captionsetup{hypcap=false}
  \caption{Comparisons for convergence rate of each tokenization method w or w/o per axis vocabulary.}
  \label{fig:exp_axis_ablation}
\end{figure}

\begin{figure}[t]
  \centering
  \includegraphics[width=\linewidth,
                   height=0.28\textheight,
                   keepaspectratio]{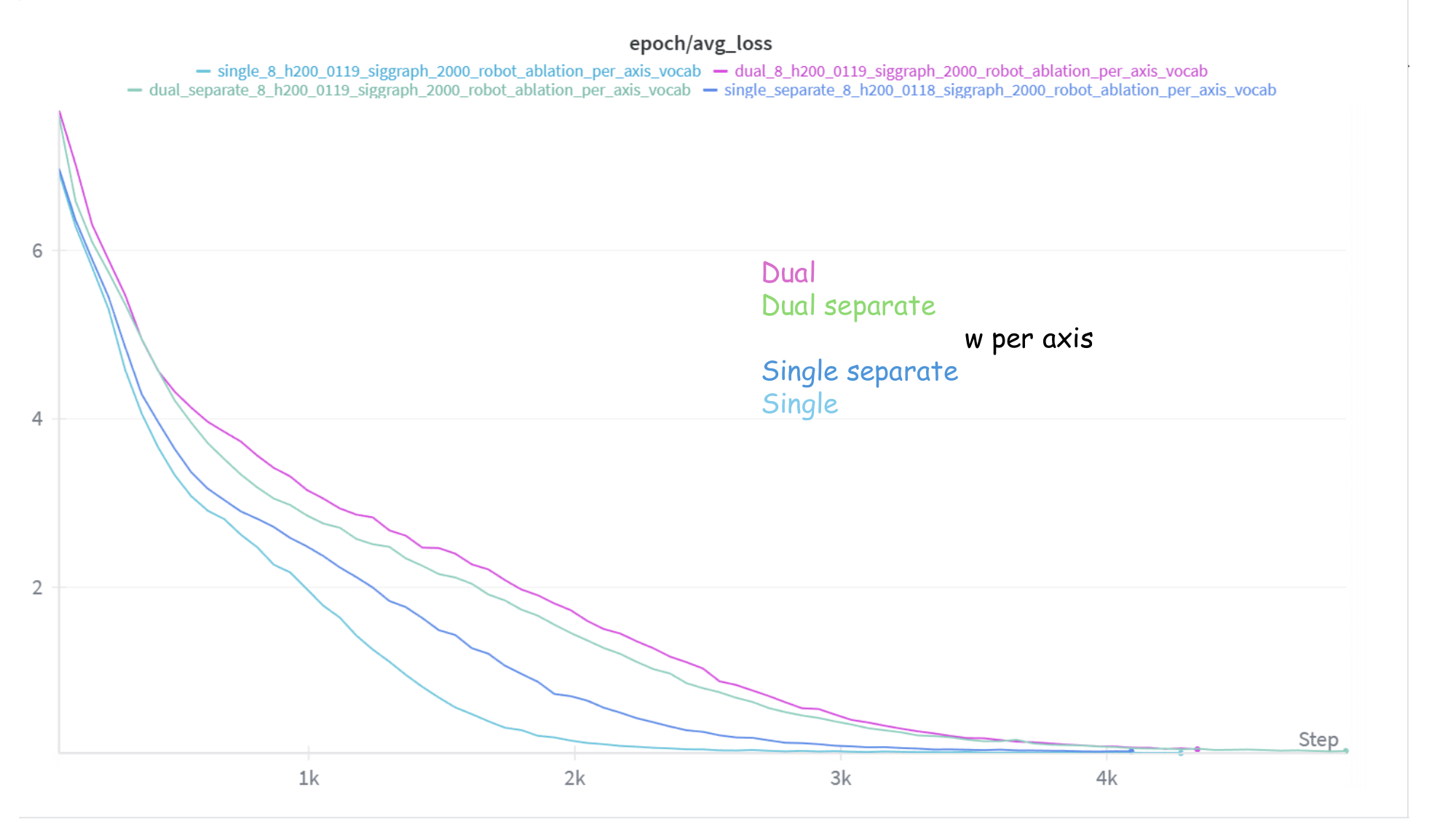}
  \captionsetup{hypcap=false}
  \caption{Comparisons for convergence rate among each tokenization method with per axis vocabulary. }
  \label{fig:exp_comparison_ablation}
\end{figure}

As shown in Fig.~\ref{fig:exp_axis_ablation}, per axis vocab has a relatively positive effect for all tokenization methods. As shown in Fig.~\ref{fig:exp_comparison_ablation}, single separate tokenization method with per axis vocabulary surpasses both dual and dual separate tokenization methods on convergence efficiency. All results are obtained on a category-balanced subset of 2,000 assets for better visualization and verified to hold on the full training set.

\subsection{Edge Flow Ratio (EFR)}
\label{sec:efr}

While CD, HD, and IoU evaluate geometric fidelity, they do not capture whether a quad-dominant mesh preserves production-ready edge flow. We therefore introduce \emph{Edge Flow Ratio} (EFR) to measure how well the edges of an output mesh align with salient feature lines of the ground-truth mesh, including sharp creases, boundary contours, and loop-like structures.

\begin{figure}[t]
  \centering
  \includegraphics[width=\linewidth,keepaspectratio]{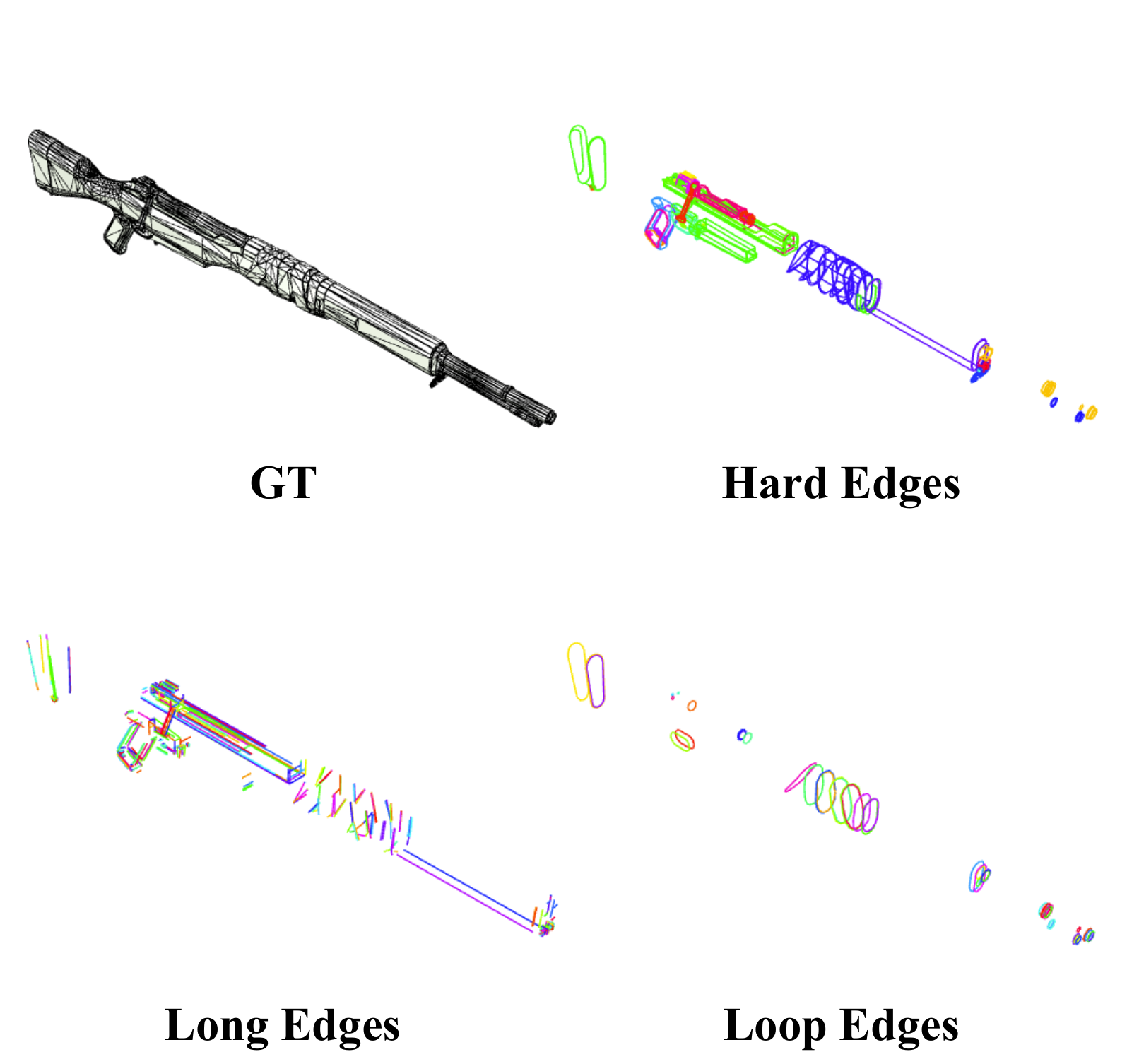}
  \captionsetup{hypcap=false}
  \caption{\textbf{Qualitative visualizations of feature line extraction for Edge Flow Ratio (EFR) calculation.}}
  \Description{Feature line extraction for Edge Flow Ratio calculation.}
  \label{fig:exp_feature_line}
\end{figure}

\paragraph{Feature Line Extraction.}
Given the ground-truth mesh $\mathcal{M}_{\mathrm{gt}}$, we first detect hard edges, including boundary edges and sharp edges whose dihedral angles exceed a threshold. Connected hard-edge chains are then grouped into two types of feature lines: long polyline features, which capture creases and ridges, and closed loop features, which capture ring-like structures such as holes or cylindrical caps. Each feature line is represented as an ordered 3D polyline. Fig.~\ref{fig:exp_feature_line} visualizes representative extracted feature lines.

\paragraph{Edge Chain Matching.}
For each ground-truth feature line $\mathbf{p}=(p_1,\ldots,p_K)$, we search for the best-matching edge chain on the output mesh $\mathcal{M}_{\mathrm{out}}$. We first resample $\mathbf{p}$ into dense points $\{\hat{p}_i\}_{i=1}^{M}$ and estimate local unit tangents $\{\hat{t}_i\}_{i=1}^{M}$. Output vertices close to the feature line are collected as
\[
\mathcal{V}_{\mathrm{near}} = \bigl\{v\in\mathcal{V}_{\mathrm{out}} \mid \min_i \|v-\hat{p}_i\|<\delta\bigr\},
\]
where each nearby vertex is associated with its closest feature-line sample
\[
\phi(v)=\operatorname*{arg\,min}_i \|v-\hat{p}_i\|.
\]

Starting from each nearby vertex, we greedily trace an edge chain by following the adjacent vertex whose outgoing direction best aligns with the local feature-line tangent:
\[
v_{k+1} = \operatorname*{arg\,min}_{u \in \mathcal{N}(v_k)\cap\mathcal{V}_{\mathrm{near}}}
\arccos\!\left(\frac{u-v_k}{\|u-v_k\|}\cdot\bigl(s\,\hat{t}_{\phi(u)}\bigr)\right),
\]
where $\mathcal{N}(v_k)$ is the one-ring neighborhood of $v_k$, and $s\in\{-1,+1\}$ resolves the traversal direction. The traversal stops when no adjacent vertex satisfies the angular threshold. We perform this search bidirectionally and retain the chain $\mathbf{q}^{*}$ with the smallest distance to $\mathbf{p}$.

\paragraph{Alignment Score and EFR.}
Given a ground-truth feature line $\mathbf{p}$ and its matched output chain $\mathbf{q}^{*}$, we uniformly resample both curves to $N_s$ points and compute the bidirectional curve distance:
\[
d(\mathbf{p},\mathbf{q}^{*}) = \frac{1}{N_s}\min\Biggl(
\sum_{i=1}^{N_s}\|\tilde{p}_i-\tilde{q}_i\|,
\sum_{i=1}^{N_s}\|\tilde{p}_i-\tilde{q}_{N_s-i+1}\|
\Biggr).
\]
The corresponding alignment score is
\[
s(\mathbf{p},\mathbf{q}^{*})
=
\exp\left(-\frac{d(\mathbf{p},\mathbf{q}^{*})}{\tau}\right),
\]
where $\tau$ controls the sensitivity to misalignment. Finally, EFR is defined as the average alignment score over all extracted feature lines:
\[
\mathrm{EFR} = \frac{1}{N_L+N_C}\Biggl(
\sum_{i=1}^{N_L}s(\ell_i,\mathbf{q}^{*}_{\ell_i})
+\sum_{j=1}^{N_C}s(c_j,\mathbf{q}^{*}_{c_j})\Biggr),
\]
where $\{\ell_i\}_{i=1}^{N_L}$ and $\{c_j\}_{j=1}^{N_C}$ denote long polyline features and loop features, respectively. Higher EFR indicates better edge-flow alignment with the ground-truth feature structure.

\paragraph{Implementation Details.}
We use proximity thresholds $\delta=0.05$ for long features and $\delta=0.01$ for loop features. The angular thresholds are set to $0.12$ radians for long features and $0.78$ radians for loops to accommodate higher curvature. We use $M=100$ samples for tangent-guided search and $N_s=500$ samples for final curve-distance computation. Feature extraction and matching are performed per connected component, with bounding-box prefiltering to skip spatially disjoint components.

\section{More Qualitative Results}

\subsection{Qualitative Comparison with Mesh Generation Methods.}
\label{app:more_qualitative_comparison}
\begin{figure*}[t]
  \centering
  \includegraphics[width=\textwidth,
                   height=0.82\textheight,
                   keepaspectratio]{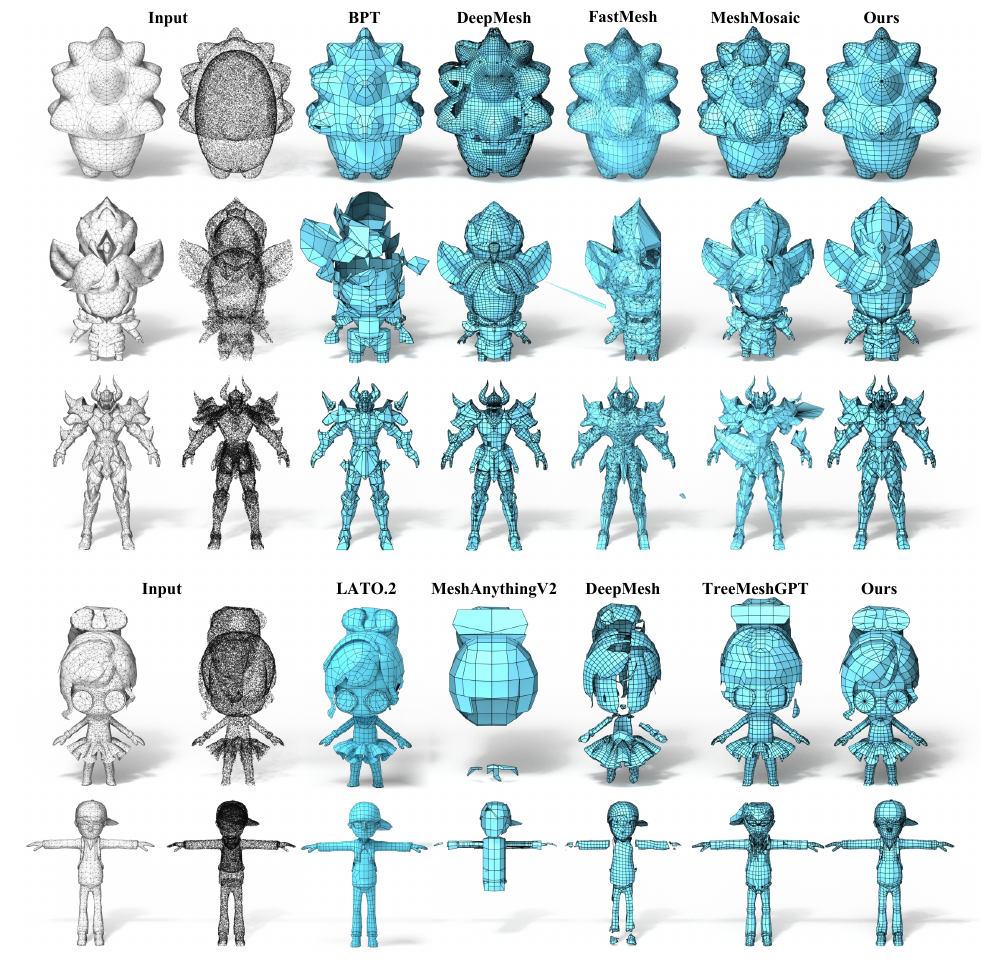}
  \caption{More qualitative comparison with triangle-based generation methods postprocessed by our \emph{Tri-to-Quad Operator}. 
  The upper three examples compare our method with BPT~\cite{weng2025scaling}, 
  DeepMesh~\cite{zhao2025deepmesh}, FastMesh~\cite{kim2025fastmesh}, 
  and MeshMosaic~\cite{xu2025meshmosaic}. 
  The lower two examples compare our method with LATO.2~\cite{long2026lato}, 
  MeshAnythingV2~\cite{chen2025meshanything}, DeepMesh~\cite{zhao2025deepmesh},
  and TreeMeshGPT~\cite{lionar2025treemeshgpt}.}
  \label{fig:app_comparison_tri}\end{figure*}

We provide more qualitative comparison with other triangle-based generation methods postprocessed by our \emph{Tri-to-Quad Operator} in Fig.~\ref{fig:app_comparison_tri}. Our method natively learns semantically structured anisotropic layouts and coherent edge flow from a dataset of quad-dominant meshes through centroid-conditioned vertex grouping, avoiding the
fragility of triangle-first generation followed by uniform geometric postprocessing.

\subsection{Field-aligned Remeshing Methods}
\label{app:quad_remeshing_comparison}

\begin{figure*}[t]
  \centering
  \includegraphics[width=\textwidth,trim=0 666.68bp 0 0,clip]{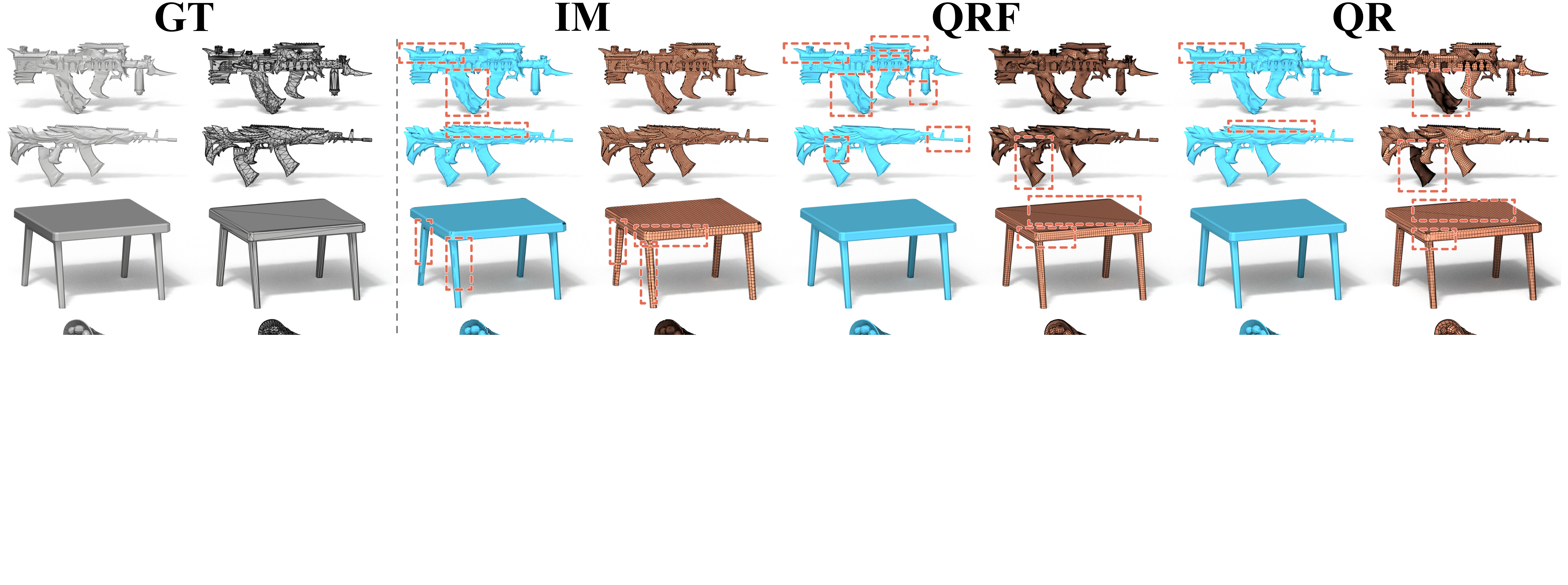}
  \caption{Qualitative visualizations of traditional quad-remeshing methods. The results show that these methods often generate unnecessarily dense face tessellations even on simple surfaces, make suboptimal topological decisions, and distort the underlying shape in pursuit of traditional mathematically regular quads. (GT = Ground‐Truth triangular mesh; IM = Instant-Meshes~\cite{jakob2015instant}; QRF = QRemeshify~\cite{qremeshify2024}; QR = Quad Remesher~\cite{quadremesher2019}.)}
  \label{fig:exp_traditional_remeshing_comparison}\end{figure*}

We present qualitative visualizations comparing with traditional quad-remeshing tools on representative meshes in Fig.~\ref{fig:exp_traditional_remeshing_comparison}. The examples highlight common failure modes of field-aligned parametrization methods, including unnecessarily dense tessellation, isotropic layouts and shape infidelity introduced to satisfy traditional mathematical quad quality criteria.

\subsection{Software-based Remeshing Baselines}
\label{app:software_remeshing_comparison}
We compare our global \emph{Tri-to-Quad Operator} against two more baselines: a greedy-based variant of our operator and a built-in algorithm in PyMeshLab~\cite{pymeshlab}. 

\noindent \textbf{Greedy Baseline.}
This baseline uses the same optimization objectives as ours, but optimizes it via local heuristics. At each step, it selects the internal edge that maximizes the immediate quality gain, updates connectivity, and repeats until no valid merges remain. Due to its purely local decisions, it frequently converges to suboptimal merge configurations, resulting in degraded quad quality.

\noindent \textbf{PyMeshLab Baseline.}
We use PyMeshLab's built-in MeshLab filter \texttt{meshing\_\allowbreak tri\_\allowbreak to\_\allowbreak quad\_\allowbreak dominant} with \texttt{level=2} (\emph{Better quad shape}), which converts a triangular mesh into a quad-dominant mesh by pairing suitable adjacent triangles. This setting is the highest-quality option provided by the filter and is used as the PyMeshLab baseline in our qualitative comparison.

\begin{figure*}[t]
  \centering
  \includegraphics[width=\textwidth]{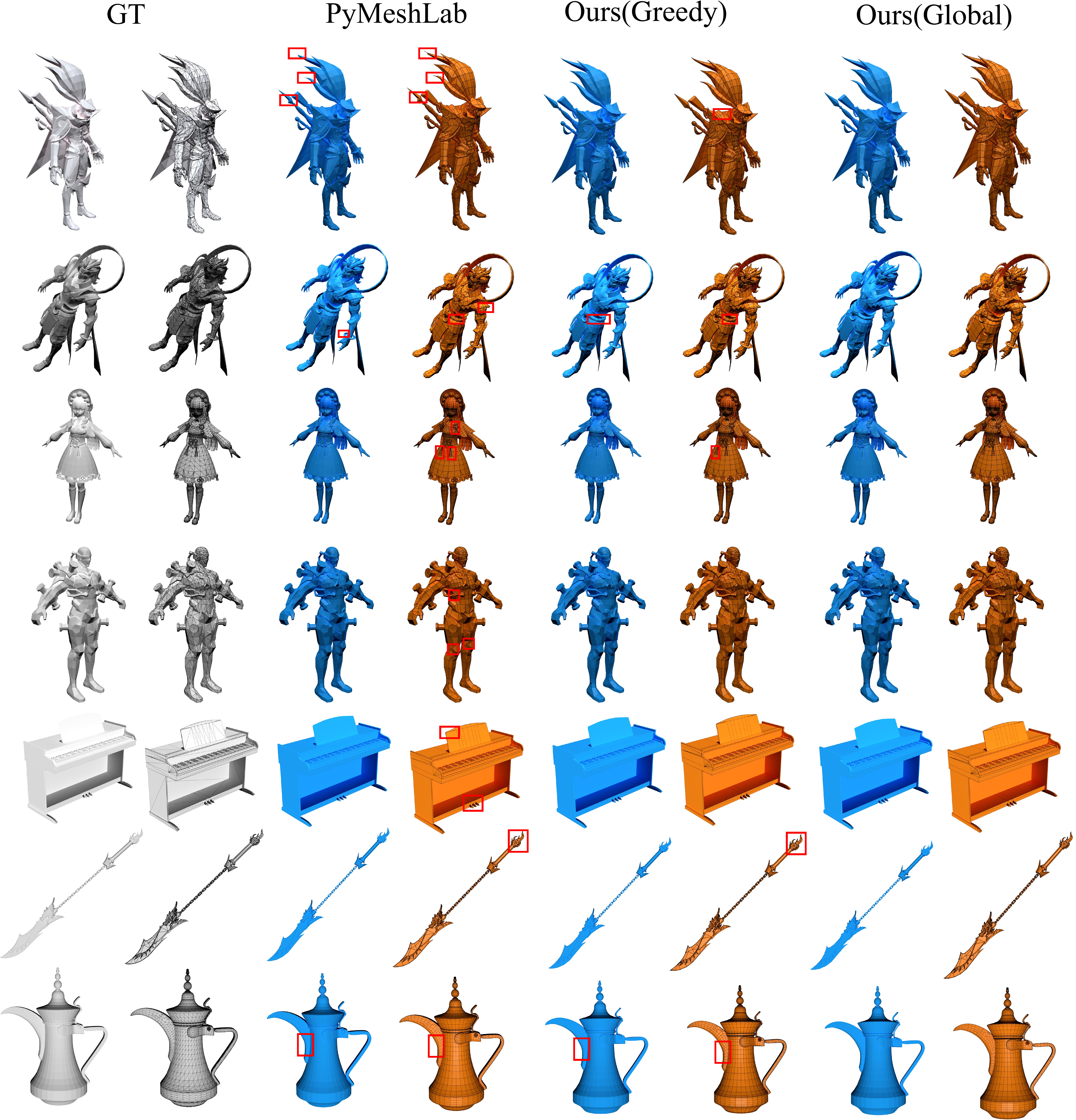}
  \caption{Qualitative comparison with Software-based
quad remeshing methods. We compare our global \emph{Tri-to-Quad Operator} against a greedy-based variant of our operator and a built-in
algorithm in PyMeshLab~\cite{pymeshlab}. Extensive results show that our operator exhibits fewer artifacts in both shape and topology.}
  \label{fig:app_operator_comparison_v3}
\end{figure*}

As illustrated in Fig.~\ref{fig:app_operator_comparison_v3}, our method produces quads that are more regular and exhibit fewer artifacts in both shape and topology under Global Merging Problem Formulation and Geometry Prefiltering.

\subsection{Polygonal Mesh Generation}
\label{app:polygonal_mesh_generation}
\begin{figure}[!htbp]
  \centering
  \includegraphics[width=\linewidth,height=0.46\textheight,keepaspectratio]{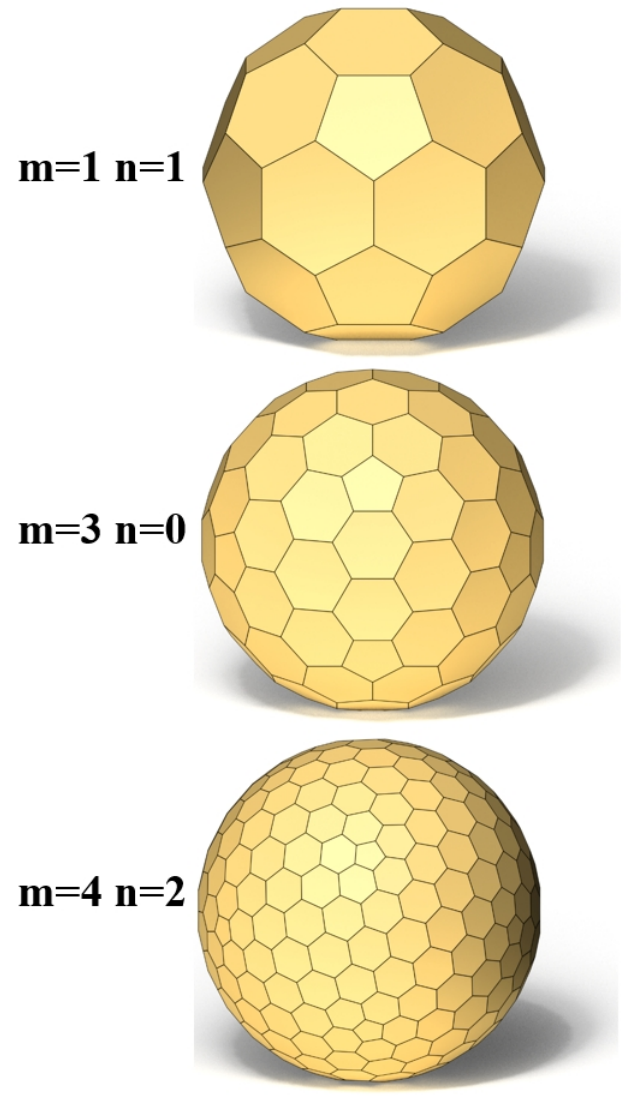}
  \captionsetup{hypcap=false}
  \caption{\textbf{Qualitative results on polygon generation with our method.}}
  \label{fig:exp_polygon}
\end{figure}

Stage~I predicts only vertices and face centroids, rather than explicit face connectivity with a fixed valence, so the same architecture can in principle represent arbitrary $n$-gon meshes. Large-scale, high-quality $n$-gon datasets are extremely difficult to obtain; we therefore use Goldberg polyhedra~\cite{liu2022extending, hart2012goldberg} to curate data and test this capability. We train on a subset of Goldberg topologies and evaluate on unseen topology values without architectural changes. As shown in Fig.~\ref{fig:exp_polygon}, QuadLink naturally generalizes to arbitrary $n$-gon meshes and generates valid Goldberg polyhedra with unseen topology values.

\paragraph{Dataset Curation.}
We procedurally construct \emph{Goldberg polyhedra} using a convex-hull-based
geodesic-dual construction. For each pair of non-negative integers $(m,n)$,
with $m \ge n$, we define the Goldberg index as
\begin{equation}
T = m^2 + mn + n^2 .
\end{equation}
We first enumerate integer lattice points $(a,b)$ inside the fundamental
Goldberg triangle,
\begin{equation}
\begin{aligned}
&(m+n)a + nb \ge 0,\quad mb - na \ge 0,\\
&T - ma - (m+n)b \ge 0.
\end{aligned}
\end{equation}
Each lattice point is mapped to the 20 faces of a unit icosahedron by
barycentric interpolation and radial projection. To avoid class-dependent
connectivity ambiguities for $n \ne 0$, we recover the geodesic triangulation
by computing the convex hull of the projected spherical point set, and then
take its topological dual to obtain the Goldberg polyhedron. The resulting
dual mesh has
\begin{equation}
\begin{aligned}
V &= 20T, \qquad E = 30T, \qquad F = 10T+2,\\
F &= 12\ \text{pentagons} + (10T-10)\ \text{hexagons}.
\end{aligned}\end{equation}

The integers $(m,n)$ are frequency parameters that control the Goldberg subdivision pattern and thereby determine the number and relative size of polygons. By varying the integer pair $(m,n)$, this procedure enables scalable generation of Goldberg meshes with controlled polygonal topology.

\paragraph{Topology Conditioning.}
As the Goldberg index $T$ increases, normalized Goldberg polyhedra become
increasingly close to the unit sphere, making their surface point clouds
nearly indistinguishable across different topologies.
This creates a \emph{degenerate conditioning} scenario: the point cloud
encoder produces nearly identical context for every Goldberg class, yet the
model must generate meshes with vastly different face counts and connectivity.
To resolve this ambiguity, we introduce a lightweight \emph{topology encoder}
that injects the Goldberg index $T$ as an explicit conditioning signal.
Concretely, a three-layer MLP maps the normalized scalar $\hat{T} = T / T_{\max}$
(with $T_{\max} = 500$) to a set of $k=4$ context tokens of dimension $d$
(matching the point cloud latent dimension):
\begin{equation}
\mathbf{c}_{\mathrm{topo}} = \mathrm{MLP}(\hat{T}) \in \mathbb{R}^{k \times d}.
\end{equation}
These tokens are concatenated to the point cloud context
$\mathbf{C}_{\mathrm{pc}} \in \mathbb{R}^{N \times d}$ before
cross-attention, yielding an augmented context
$\mathbf{C} = [\mathbf{C}_{\mathrm{pc}};\, \mathbf{c}_{\mathrm{topo}}]
\in \mathbb{R}^{(N+k) \times d}$.
During training, $T$ is derived from the sample identifier. During inference, the user specifies the desired $T$ to steer generation toward the
target topology. The topology encoder adds only ${\sim}0.5$K learnable
parameters and introduces no architectural change to the mesh transformer
itself.

\paragraph{Experiments and Results.}
We enumerate valid $(m,n)$ pairs with unique $T$ values and select 100
topologically distinct Goldberg classes, spanning $T \in [1,324]$ and face
counts from 12 to 3242. For each topology, we generate one canonical mesh and four randomly rotated variants, yielding 500 meshes in total. All meshes are normalized to $[-1,1]^3$. We use 95 topologies for training and reserve 5 unseen topologies for testing.

We finetune our pretrained model without architectural changes. The only
modification is in Face Assembly \textit{(Stage~III)}, where we replace the
\emph{quad-first, tri-next} strategy with a \emph{hexagon-first, pentagon-next} strategy.

\subsection{Efficiency/Token Length Analysis.}
\label{app:efficiency_token_length}

Training and inference times in autoregressive mesh generation are influenced by generation completeness, model parameters, data distribution, sliding-window size and many other factors, making direct wall-clock comparisons unfair. Therefore, we follow common practice by using token length to reflect the representation-level efficiency gain, since dense self-attention scales as $\mathcal{O}(N^2)$ in sequence length.

For Meshtron's three-level face--vertex--coordinate tokenization~\cite{hao2024meshtron} (Sec.~2), each triangular face contributes three vertices with three coordinates each, giving

\begin{equation}
N_{\mathrm{Meshtron}} = 3 \times 3 \times F_{\triangle} = 9F_{\triangle}.
\end{equation}
Because vertices shared by adjacent faces are repeatedly tokenized, this representation scales with the number of triangle faces. QuadLink instead predicts each unique vertex and face center once, giving
\begin{equation}
N_{\mathrm{ours}} \approx 3\bigl(V + F_{\mathrm{poly}}\bigr).
\end{equation}

For a closed, genus-zero triangular mesh, Euler's formula $V-E+F_{\triangle}=2$ and the face--edge incidence relation $3F_{\triangle}=2E$ imply $F_{\triangle}=2V-4\approx2V$, so $N_{\mathrm{Meshtron}}=9F_{\triangle}\approx18V$. For a closed genus-zero mesh containing only triangles and quads, define $q=F_4/F_{\mathrm{poly}}$ as the fraction of quad faces, where $F_3=(1-q)F_{\mathrm{poly}}$ and $F_4=qF_{\mathrm{poly}}$. Since every edge is incident to two faces, $2E=3F_3+4F_4=(3+q)F_{\mathrm{poly}}$. Substituting $E=(3+q)F_{\mathrm{poly}}/2$ into Euler's formula $V-E+F_{\mathrm{poly}}=2$ gives

\begin{equation}
F_{\mathrm{poly}} = \frac{2(V-2)}{1+q} \approx \frac{2V}{1+q}.
\end{equation}

QuadGPT~\cite{liu2026quadgpt} assigns a fixed 12-token block to each face: 12 coordinate tokens for a quad and nine coordinate tokens plus three padding tokens for a triangle. With $N_{\mathrm{ours}}\approx3(V+F_{\mathrm{poly}})$, the token-count ratios are
\begin{equation}
\begin{aligned}
\frac{N_{\mathrm{ours}}}{N_{\mathrm{Meshtron}}} &\approx \frac{3(V+F_{\mathrm{poly}})}{18V} \approx \frac{q+3}{6(q+1)},\\
N_{\mathrm{QuadGPT}} &= 12F_{\mathrm{poly}},\\
\frac{N_{\mathrm{ours}}}{N_{\mathrm{QuadGPT}}} &\approx \frac{3(V+F_{\mathrm{poly}})}{12F_{\mathrm{poly}}} \approx \frac{q+3}{8}.
\end{aligned}
\end{equation}

For a pure triangle mesh ($q=0$), QuadLink uses about half as many tokens as Meshtron. For a pure quad mesh ($q=1$), it uses about one-third as many. Mesh-Pro~\cite{zhou2026mesh} also uses 12 tokens per face. Compared with QuadGPT and Mesh-Pro, QuadLink uses 37.5\%--50\% as many tokens, depending on the quad fraction $q$.

DeepMesh and BPT shorten mesh sequences with block-wise or patch-wise tokenization~\cite{zhao2025deepmesh,weng2025scaling}. This compression trades sequence length against vocabulary size and structure, while making the face--vertex--coordinate hierarchy less explicit. If generation stops before a block or patch is complete, local face connectivity may remain unresolved, increasing the risk of holes.

\subsection{Ablation on the Retrieval Space}
\label{app:retrieval_space_ablation}

For each face centroid, Link Modeling (Stage II) ranks vertices by their distance to the centroid in a retrieval space. Face Assembly (Stage III) then tests combinations from the nearest candidates using Geometry Prefiltering and Centroid Tolerance. The retrieval space determines which vertices are considered, and the geometric checks determine which combinations form valid faces.

\noindent\textbf{Retrieval spaces and supervision.}
We compare Euclidean distance in the original coordinates, distance between hyperspherical embeddings which is the latent-space retrieval in our QuadLink and the spacetime interval used for vertex adjacency in SpaceMesh~\cite{shen2024spacemesh} and Nexus~\cite{wang2026nexus}. We adapt the interval to centroid--vertex retrieval. For learned embeddings, threshold-based binary cross-entropy (BCE) supervises each centroid--vertex pair by its incidence label used in SpaceMesh~\cite{shen2024spacemesh} and Nexus~\cite{wang2026nexus}. Triplet margin loss used in our QuadLink instead requires an incident vertex to be closer to the centroid than a non-incident vertex by a margin.

Exp.~0 is the Euclidean baseline and requires no training. Exp.~1 and Exp.~2 compare hyperspherical and spacetime retrieval with BCE. Exp.~3 uses a bounded spacetime interval with triplet margin loss, and Exp.~4 is our hyperspherical triplet model. Tab.~\ref{tab:retrieval_space_settings} lists the distances and training parameters. The spacetime comparison between Exp.~2 and Exp.~3 changes both supervision and embedding normalization.

\begin{table}[t]
    \centering
    \footnotesize
    \setlength{\tabcolsep}{2.5pt}
    \caption{\textbf{Retrieval spaces and supervision for the Stage II ablation.}}
    \label{tab:retrieval_space_settings}
    \begin{tabular}{@{}llll@{}}
        \toprule
        Exp. & Name & Distance $d$ & Objective and parameters \\
        \midrule
        0 & 3D Euclidean & $\lVert x_c-x_v\rVert_2$ & No training \\
        \addlinespace[2pt]
        1 & Hypersphere + BCE & $\lVert \hat z_c-\hat z_v\rVert_2^2$ & \begin{tabular}[t]{@{}l@{}}BCE: $\ell=\alpha(\tau-d)$ \\ Init.: $\alpha=10$, $\tau=2$\end{tabular} \\
        \addlinespace[2pt]
        2 & \begin{tabular}[t]{@{}l@{}}Spacetime + BCE \\ (Nexus-style)\end{tabular} & $d_{\mathrm{st}}(s,t)$ & \begin{tabular}[t]{@{}l@{}}BCE: $\ell=\tau-d$ \\ Init.: $\tau=0$\end{tabular} \\
        \addlinespace[2pt]
        3 & \begin{tabular}[t]{@{}l@{}}Bounded Spacetime \\ + Triplet\end{tabular} & $\lambda d_{\mathrm{st}}(\hat s,\hat t)$ & Triplet, $\lambda=0.25$ \\
        \addlinespace[2pt]
        4 & \begin{tabular}[t]{@{}l@{}}Hypersphere + Triplet \\ (Ours)\end{tabular} & $\lVert \hat z_c-\hat z_v\rVert_2$ & Triplet \\
        \midrule
        \multicolumn{4}{@{}p{\columnwidth}@{}}{For centroid $c$ and vertex $v$, $x$ denotes the coordinates, $z=[s,t]$ the embedding, and a hat denotes unit normalization. The interval is $d_{\mathrm{st}}(s,t)=\lVert s_c-s_v\rVert_2^2-\lVert t_c-t_v\rVert_2^2$.} \\
        \addlinespace[2pt]
        \multicolumn{4}{@{}p{\columnwidth}@{}}{BCE uses a learned threshold $\tau$, with a learned scale $\alpha$ in Exp.~1. Both variants use the four-group balancing of Nexus. For Exp.~2, the embedding is scaled by $(D/2)^{-1/4}$ without unit normalization, where $D$ is dimension.} \\
        \addlinespace[2pt]
        \multicolumn{4}{@{}p{\columnwidth}@{}}{Both triplet variants use online hard negative mining (Sec.~\ref{app:hard_mining}), with margin $m$: $0.2\to 0.3$ and mining budget $k_{\mathrm{hard}}$: $20\to 50$.} \\
        \bottomrule
    \end{tabular}
\end{table}

\noindent\textbf{Evaluation.}
The learned variants share a cross-attention backbone, initialization, and optimization schedule. We train for 80 epochs on 3,000 gun asset groups and select checkpoints by validation retrieval. We evaluate on 290 held-out meshes from the same category, covering 629,875 faces. To isolate retrieval and assembly, we use ground-truth vertices and face centroids. For each centroid, all variants rank candidates by distance and pass the top $K$ to the same Face Assembly procedure. We use $K=20$ and $K=6$ to measure how each ranking performs under a tighter candidate budget.

\begin{table}[t]
    \centering
    \small
    \setlength{\tabcolsep}{2.5pt}
    \caption{\textbf{Quantitative ablation on Stage II vertex retrieval under the oracle setting.} Recall is reported in percent. The best results are in bold.}
    \label{tab:retrieval_space_ablation}
    \begin{tabular}{@{}lrrrrrr@{}}
        \toprule
        Exp. & R@4$\uparrow$ & R@6$\uparrow$ & R@12$\uparrow$ & R@20$\uparrow$ & mAP$\uparrow$ & MWPR$\downarrow$ \\
        \midrule
        0 & \textbf{12.38} & 29.88 & 64.28 & 75.99 & 0.4974 & 49.79 \\
        1 & 6.65 & 20.02 & 55.51 & 74.94 & 0.4878 & \textbf{20.26} \\
        2 & 5.96 & 17.84 & 50.15 & 68.57 & 0.4492 & 26.61 \\
        3 & 12.25 & \textbf{31.87} & \textbf{74.13} & \textbf{87.00} & 0.5790 & 74.01 \\
        4 (Ours) & 12.01 & 31.41 & 73.53 & 86.91 & \textbf{0.5797} & 20.98 \\
        \bottomrule
    \end{tabular}
\end{table}

\begin{table}[t]
    \centering
    \small
    \setlength{\tabcolsep}{5pt}
    \caption{\textbf{Quantitative ablation on Stage II vertex retrieval under high valence and 3D crowding.} Recall is reported in percent.}
    \label{tab:retrieval_hard_cases}
    \begin{tabular}{@{}lrrr@{}}
        \toprule
        & High valence & \multicolumn{2}{c@{}}{High 3D crowding} \\
        \cmidrule(lr){2-2}\cmidrule(l){3-4}
        Exp. & R@4$\uparrow$ & R@4$\uparrow$ & R@6$\uparrow$ \\
        \midrule
        0 & 6.40 & $\approx 0$ & 1.31 \\
        1 & 6.98 & 0.89 & 4.68 \\
        2 & 6.49 & 0.90 & 4.54 \\
        3 & \textbf{13.41} & 2.41 & 10.75 \\
        4 (Ours) & 13.38 & \textbf{2.47} & \textbf{11.05} \\
        \bottomrule
    \end{tabular}
\end{table}

\begin{table}[t]
    \centering
    \small
    \setlength{\tabcolsep}{2.5pt}
    \caption{\textbf{Quantitative ablation on Face Assembly (Stage III) under two oracle candidate budgets.} Coverage, exact match, and non-manifold edge ratio (NME) are in percent. Time is the mean assembly time per mesh in seconds.}
    \label{tab:retrieval_space_reconstruction}
    \begin{tabular}{@{}lrrrrrr@{}}
        \toprule
        Exp. & Cov.$\uparrow$ & Match$\uparrow$ & Face F1$\uparrow$ & Edge F1$\uparrow$ & NME$\downarrow$ & Time$\downarrow$ \\
        \midrule
        \multicolumn{7}{@{}l}{Oracle ($K=20$)} \\
        0 & 86.44 & 27.73 & 0.3015 & 0.4652 & 5.25 & \textbf{19.50} \\
        1 & 92.61 & 24.84 & 0.2634 & 0.4333 & 5.73 & 22.32 \\
        2 & 88.15 & 22.90 & 0.2487 & 0.4155 & 5.80 & 20.64 \\
        3 & 96.96 & \textbf{29.79} & \textbf{0.3074} & \textbf{0.4723} & 5.29 & 21.98 \\
        4 (Ours) & \textbf{97.09} & 29.74 & 0.3065 & 0.4712 & \textbf{5.11} & 22.10 \\
        \midrule
        \multicolumn{7}{@{}l}{Oracle ($K=6$)} \\
        0 & 58.98 & 18.79 & 0.2405 & 0.3876 & 3.83 & 1.47 \\
        1 & 54.34 & 12.85 & 0.1721 & 0.3268 & 4.83 & 1.25 \\
        2 & 50.76 & 11.42 & 0.1562 & 0.2982 & 4.83 & \textbf{1.20} \\
        3 & 66.64 & \textbf{19.84} & \textbf{0.2429} & \textbf{0.3997} & 3.34 & 1.62 \\
        4 (Ours) & \textbf{67.29} & 19.60 & 0.2388 & 0.3971 & \textbf{3.18} & 1.62 \\
        \bottomrule
    \end{tabular}
\end{table}

\noindent\textbf{Results.}
\textit{Vertex retrieval.} Tab.~\ref{tab:retrieval_space_ablation} evaluates the ranking of incident vertices. R@$k$ is the fraction of faces with all ground-truth vertices in the top $k$. Mean average precision (mAP) averages precision at the ranks of incident vertices, first within each face and then across faces. Mean worst positive rank (MWPR) averages the rank of the last incident vertex. The triplet variants outperform the BCE variants. Under triplet supervision, the two retrieval spaces yield similar recall and mAP, with a lower MWPR on the hypersphere.

\noindent\textit{Hard cases.} Tab.~\ref{tab:retrieval_hard_cases} measures R@$k$ on two difficult subsets. A high-valence face touches a vertex shared by at least seven faces. A crowded face has at least four non-incident vertices within the centroid's distance to its farthest incident vertex. Both triplet variants improve recall over Euclidean retrieval in these subsets, with the largest gains in crowded neighborhoods.

\noindent\textit{Face assembly.} Tab.~\ref{tab:retrieval_space_reconstruction} tests how the rankings affect assembly under two candidate budgets. Coverage and exact match are the fractions of centroids that yield a valid face and the exact reference vertex set, respectively. Face F1 and Edge F1 summarize precision and recall over the recovered faces and edges. NME is the fraction of predicted edges shared by more than two faces, and Time is the mean assembly time per mesh. At both budgets, the triplet variants achieve higher Face and Edge F1 than the BCE variants. The bounded spacetime variant has slightly higher F1 than ours, while ours has higher coverage and a lower NME. A tighter budget speeds up assembly at the cost of topology recovery.

\noindent\textbf{Efficiency.}
In the Face Assembly ablation in Tab.~\ref{tab:ablation_stage2}, latent-space retrieval reduces reconstruction time with both Geometry Prefiltering and Centroid Tolerance enabled. To further accelerate Face Assembly, we cache candidate combination indices and skip combinations already checked in smaller pools. For each centroid, we evaluate the centroid tolerance of candidate combinations in batches and apply the remaining geometric checks to the surviving combinations in their original order. The search retains the quad-first selection rule and falls back to triangles when no valid quadrilateral is found. We then compute the distances and stable vertex ranking once. In a matched check on 24 faces from one mesh with 3D retrieval, this reduces assembly time from $17.22$ to $0.385$\,s at $K=20$ and from $0.249$ to $0.183$\,s at $K=6$, with identical selected faces and output ordering. Across all 290 meshes, Exp.~4 averages $22.10$\,s per mesh at $K=20$ and $1.62$\,s at $K=6$.

\noindent\textbf{Choice of retrieval space.} The two triplet variants achieve comparable retrieval and assembly quality, with different trade-offs in topology and coverage in Tab.~\ref{tab:retrieval_space_reconstruction}. At nearly identical R@20, hyperspherical retrieval yields a substantially lower MWPR in Tab.~\ref{tab:retrieval_space_ablation}, indicating that when the top 20 candidates miss an incident vertex, the last incident vertex tends to rank less far down the list. Based on this ranking behavior, we adopt hyperspherical retrieval with triplet margin loss for large-scale training.

\section{Discussions}
\label{sec:appendix_discussions}

\subsection{Singularities, Watertightness and Manifoldness}
\label{app:topology_discussion}

There remains a substantial gap between the geometric criteria commonly emphasized in traditional remeshing pipelines and the practical requirements of production-ready mesh design or generation. Traditional geometry processing methods often prioritize mathematical regularity, such as minimizing singularities, enforcing watertightness, and preserving manifoldness. In contrast, production-ready assets are frequently optimized for editability and compatibility with downstream workflows, where such criteria are not always strictly enforced and may even be intentionally relaxed.

\noindent\textbf{Singularities.}
In a pure quadrilateral mesh, a regular interior vertex is expected to have valence four, which indicates it is incident to four quad edges. Vertices with non-four valence, such as valence three, five, or higher, are commonly referred to as singularities. In traditional quad remeshing, singularities play a central role because they determine how the global edge flow branches, merges, or changes direction. Consequently, traditional field-aligned pipelines often aim to minimize the number of singularities, place them at geometrically meaningful locations, and maintain smooth cross field orientation with near-uniform quad sizes.

However, this notion is less directly aligned with production workflows. Game and animation assets are rarely required to be mathematically clean pure-quad meshes throughout the entire pipeline. Instead, artists commonly work with quad-dominant meshes during modeling and editing, while final assets may be triangulated for rendering or engine deployment. Once triangles and occasional $n$-gons are introduced, strict valence-based singularities become abundant and less informative as a standalone quality measure. In such settings, production quality is more often determined by whether the mesh exhibits coherent and controllable edge flow. Well-structured edge flow supports common artist operations, including UV unwrapping, beveling, subdivision, and local editing, even when the mesh contains singularities or mixed face types. Therefore, rather than treating singularities as artifacts to be universally minimized, production-oriented topology often uses them as localized flow-control mechanisms that help redirect, terminate, or connect edge loops in a practical and editable manner.

\noindent\textbf{Watertightness and Manifoldness.}
Watertightness and manifoldness are also central criteria in traditional geometry processing, especially for simulation, physical analysis and volumetric processing, where closed and well-defined surface topology is often required. A watertight mesh forms a closed surface without holes or boundary gaps, while a manifold mesh ensures that the local neighborhood of each point behaves like a disk or half-disk. These properties make the surface mathematically well behaved and simplify many downstream algorithms.

However, production-ready assets are not always optimized under these strict assumptions. In game, animation and content-creation pipelines, meshes are often composed of multiple disconnected components, overlapping parts, open surfaces, clothing or hair layers, accessories, thin shells for preserving well-designed shape details and topological editing flexibility. Such structures may be non-watertight or locally non-manifold, but they are still valid and useful production assets as long as they support operations such as modeling, editing, texturing, rigging, rendering, asset assembly and so on. 

Therefore, while watertightness and manifoldness remain important for specific applications such as simulation or manufacturing, they are not uniform indicators for all production pipelines. For editable asset generation, overly enforcing these constraints may remove semantic structural separation, alter part boundaries, or introduce unnecessary geometric and topological repairs. In this work, we therefore focus on generating quad-dominant meshes that better support production workflows for games and animation, rather than enforcing watertight or manifold structure as strict objectives.\subsection{Topology Preservation and Layout Diversity}
\label{app:topology_preservation}

Current autoregressive methods lack strict mathematical guarantees to preserve the macro-topology (e.g., genus) implied by the point cloud, although they empirically achieve high topological fidelity. For micro-topology (mesh connectivity), a single surface naturally admits multiple valid quad layouts. Our model embraces this one-to-many mapping by adjusting sampling parameters ($T$, top-$k$, and top-$p$), enabling diverse yet plausible generated layouts for the same input. Incorporating explicit topological validation remains an important future direction for exploration.

\subsection{Limitations}
\label{app:limitations}

While QuadLink learns directly from production-ready quad-dominant meshes with semantically anisotropic layouts and coherent edge flow, its Anchor Prediction \textit{(Stage I)} still follows an autoregressive generation paradigm. As a result, QuadLink shares several limitations commonly observed in autoregressive 3D generative models, especially in terms of interactive and controllable remeshing. In traditional field-aligned parametrization pipelines such as Instant-Meshes~\cite{jakob2015instant}, users can often provide interactive guidance, such as symmetry constraints and target polygon counts. In contrast, integrating such controls into an autoregressive mesh generative model remains non-trivial, because these constraints are not naturally represented as the same type of coordinate tokens used for point prediction.

\noindent\textbf{Symmetry Control.}
For symmetric shapes, users often expect the generated topology to preserve symmetry across corresponding parts. For example, if an input character has two approximately symmetric arms, the remeshing results should ideally maintain symmetric topology on both sides. However, small geometric asymmetries in the input point cloud, such as an accessory attached to only one arm, can perturb the autoregressive generation process. Since QuadLink orders anchor tokens according to a fixed spatial ordering and predicts them sequentially, such local asymmetries may propagate through next-token prediction and lead to asymmetric topology on regions that are semantically expected to remain symmetric. One possible direction~\cite{zhou2026quartet} is to explicitly generate or condition on symmetry axes before generation. However, introducing such structural tokens increases sequence length and creates a mixed-token representation whose semantics differ from standard mesh coordinate tokens, potentially making autoregressive training less stable.

\noindent\textbf{Target Polygon Count.}
Another limitation is explicit control over the target polygon count. In production pipelines, artists often require different levels of detail (LODs), where each asset must satisfy a target polygon count. Current autoregressive mesh generative models cannot reliably control vertex or face counts by conditioning on a scalar target count. This limits their direct use in workflows that require strict polygon budgets. Future work may require stronger control mechanisms or alternative generative formulations, such as discrete diffusion or refinement-based architectures~\cite{song2025topology}, to support controllable polygon-count generation.

Overall, traditional field-aligned parametrization methods are easier to combine with interactive controls because such constraints can be directly written into rule-based optimization pipelines. However, their reliance on handcrafted rules also limits their robustness and generalization to complex production assets. In contrast, 3D generative models such as QuadLink provide stronger data-driven modeling capacity, but still lack a unified representation for interactive constraints. We believe that future modular designs or hybrid generative architectures that explicitly encode symmetry, polygon counts and other interactive information will lead to fully controllable production-ready mesh generation.

\par\bigskip
\begingroup
\small
\bibliographystyle{ieeenat_fullname}
\bibliography{references}
\endgroup

\end{document}